\documentclass[aps,prb,reprint,groupedaddress, showpacs]{revtex4-1}
\usepackage{graphicx,graphics}
\usepackage{epstopdf}
\usepackage{dcolumn}
\usepackage{amsmath,amssymb,amsfonts}
\usepackage{latexsym,verbatim}
\usepackage{bm}
\usepackage{bbold}
\usepackage{color}
\usepackage{ulem}
\usepackage[breaklinks=false,colorlinks,citecolor=blue,linkcolor=blue,urlcolor=blue]{hyperref}
\usepackage[abs]{overpic}
\begin{document}
\title{Adiabatic perturbation theory of nonequilibrium light-controlled superconductivity}
\author{Andrea Secchi}
\email{Andrea.Secchi@iit.it}
\affiliation{Istituto Italiano di Tecnologia, Graphene Labs, Via Morego 30, I-16163 Genova, Italy}
\author{Marco Polini}
\affiliation{Istituto Italiano di Tecnologia, Graphene Labs, Via Morego 30, I-16163 Genova, Italy}
\begin{abstract}
Recent experiments, in which Terahertz (THz) light has been used to induce nonequilibrium superconducting states, have raised a number of intriguing fundamental questions. Theoretically, these experiments are most often described within the Floquet formalism, which suffers a number of well-known limitations (e.g.~Floquet heating). Alternative approaches rely on heavy numerical methods. In this Article we develop an analytical theory of nonequilibrium superconductivity that combines path integrals on the Kostantinov-Perel' time contour with adiabatic perturbation theory [G. Rigolin, G. Ortiz, and V.H. Ponce, Phys. Rev. A~{\bf 78}, 052508 (2008)]. We consider a general system of electrons and Raman phonons coupled by the Fr\"ohlich interaction, in the presence of a time-dependent external field which acts on the phonon subsystem. The latter is supposed to model the THz light-induced excitation of nonlinear interactions between infrared and Raman phonons. Assuming that the external field has a slow dependence on time, we derive equations for the dynamical superconducting gap, calculating the leading adiabatic term and the first non-adiabatic correction. Our nonequilibrium formulas can be solved numerically with a minimal increase of computational complexity with respect to that needed to calculate the superconducting gap at equilibrium.
\end{abstract}

\date{\today}

\maketitle

\section{Introduction}
\label{sec: Intro}

Discovering or engineering materials displaying superconductivity at room temperature represents an extraordinary challenge, with obvious disruptive technological implications. Since the critical temperature $T_{\rm c}$ for conventional BCS superconductors
is typically as low~\cite{Tinkham} as $\approx  10^{0} \sim 10^{1}~{\rm K}$, large theoretical and experimental efforts are being devoted to the search for high-temperature superconductors~\cite{Anderson13, Keimer15} and generalizations of the equilibrium theory beyond the Eliashberg equations~\cite{Pietronero95, Grimaldi95}. 

On the other hand, recent advances in the production and manipulation of intense Terahertz light sources have triggered a very interesting question. Is it possible to turn a normal material into a superconducting one, at least {\it temporarily},  by applying an appropriately designed time-dependent electromagnetic field? More precisely, recent experiments indicate that stimulation by light of a superconducting material at temperatures {\it above} $T_{\rm c}$, even up to room temperature, may induce in the otherwise normal state at least some of the properties of the superconducting phase (e.g.~coherent transport), avoiding the need to cool the material down to very low temperatures~\cite{Fausti11, Hu14, Kaiser14, mitrano_nature_2016}. Of course this approach costs energy. For any technological application one should therefore assess whether the pros of operating a room-temperature nonequilibrium superconducting phase overcome the cons linked to sustaining the electromagnetic field over a certain time window. Ignoring such practical considerations, this fascinating question challenges our understanding of the mechanisms of interaction between THz light, phonons, and electrons.

From the point of view of theory, the main goal is to quantify how the superconducting gap ($\Delta$) changes in time due to the presence of an external time-dependent field. General integro-differential equations have been derived, e.g.~within the formalism of Keldysh nonequilibrium Green's functions~\cite{RammerSmith, Kopnin}, for the Bardeen-Cooper-Schrieffer (BCS) time-dependent gap. Despite their generality, their solutions rely on approximations or assumptions that limit their applicability (see e.g.~the case of a dirty superconductor~\cite{Usadel, RammerSmith, Kamenev, Semenov16}). One can group different theoretical approaches to nonequilibrium superconductivity on the basis of the time dependence of the external field. Three cases occur: (1) the field changes {\it slowly}, or (2) {\it quickly}, on the characteristic time scales set by the equilibrium parameters of the system (a condition that can be specified in different ways~\cite{Kopnin, Abrahams66}), and/or (3) the field is {\it periodic} in time. In our case, we say that the external field is slow if the transition amplitude between instantaneous eigenstates of the Hamiltonian $\hat{\mathcal{H}}(t)$ induced by its time-derivative is much smaller than the ratio between the energy gap between those states and the time scale over which the system is observed ($T$). This requirement on the smallness of the time-derivative of the external field will remind the reader of the conditions of validity of the adiabatic theorem in quantum mechanics~\cite{Messiah}. Below we will see that we actually need a more powerful formalism. With this definition of the rapidity of variation of the external field, the approach that we pursue in this work deals with problems belonging to group (1). Let us briefly comment on other approaches first.

In relation to problems of type (2), many theoretical works have focussed on the non-adiabatic regime, which often requires a fully numerical treatment. Typical problems that have been investigated in the literature involve: i) an instantaneous switching-on of the field~\cite{Barankov04, Warner05}; ii) a quench of the attractive interaction between anti-parallel spin electrons~\cite{Volkov74, Yuzbashian06, Barankov06, Tomadin08, Peronaci15}; iii) an ultrafast but non-instantaneous (e.g.~Gaussian) external field acting on the electron subsystem~\cite{Papenkort07, Unterhinninghofen08}; iv) simulations of ultrafast pump-probe experiments in superconductors~\cite{Krull14}; v) preparation of the system in a nonequilibrium state and study of its evolution under a time-independent Hamiltonian~\cite{Yuzbashyan05}. The motivation to study this regime is given by new experimental techniques for the ultrafast optical manipulation of superconductivity, including real-time tracking of the evolution of $\Delta$~\cite{Mansart13,Matsunaga14}.

When the external field is {\it periodic} in time, as in the case of problems of type (3), the Floquet formalism provides the simplest way to compute time-dependent observables. In Ref.~\onlinecite{Coulthard16}, this approach was applied to the Hubbard model, showing that the super-exchange interaction can be modulated to become the dominant energy scale of the system, switching-on pair correlations that are responsible for superconductivity. In Ref.~\onlinecite{Knap16}, Floquet theory was used to analyze Cooper-pair instabilities in nonequilibrium electron-phonon systems. The effective electron-electron interaction resulting from the electron-phonon coupling was treated in the Hubbard approximation and a quartic time-periodic phonon driving effectively modified the interaction parameter in time. During the transient, at low driving frequencies, a competition takes place between Cooper-pair enhancement due to the driving and Cooper-pair breaking due to the nonequilibrium distribution of phonons. The critical temperature $T_{\rm c}$, defined with respect to the time-averaged Hamiltonian, was found to increase in a broad region of parameter space, with a complicated dependence on the driving frequency. 

Despite its usefulness, the Floquet formalism can be strictly applied only when the driving is perfectly periodic, which is not consistent with realistic experimental situations. Establishing a field that can be modelled as periodic requires a switching-on procedure occurring on a long time scale, which may cost a significant amount of energy. For a rigorous application of Floquet theory, the field should then last forever. Moreover, a well-known problem with Floquet theory is the phenomenon of {\it Floquet heating}, by which an interacting system heats up to an infinite temperature at infinite times. Although in some situations heating is slow enough in the time interval of interest~\cite{Knap16, Bukov16}, it is nevertheless a non-physical effect whose impact needs to be carefully addressed. Recently, the authors of Ref.~\onlinecite{Kuwahara16} have demonstrated that, under rather general conditions, the transient dynamics of Floquet systems, on a finite time scale $\tau^*$, can be accurately described by means of the high-frequency Floquet-Magnus expansion truncated at a certain optimal order, which depends on $\tau^*$.

The limitations of Floquet theory can be overcome via fully numerical approaches. For example, the authors of Ref.~\onlinecite{Sentef16} investigated the nonequilibrium dynamics of a phonon-mediated superconductor induced by a transiently modified electronic structure through nonlinear phonon coupling. The system was modelled by a Fr\"ohlich Hamiltonian with a dynamical electronic band structure (i.e.~a two-dimensional square-lattice tight-binding model with time-dependent hopping). 
The time-dependent hopping amplitude was taken to evolve linearly with time from an initial value $J_0 = 0.25~{\rm eV}$ to a final value $J_{\rm f} = 0.20~{\rm eV}$ (reached after a time $\tau$) and stayed equal to $J_{\rm f}$ afterwards. For this model, the Kadanoff-Baym equations for the nonequilibrium Green's functions were solved numerically. The authors demonstrated an enhancement of the superconducting gap and discussed mechanisms and time scales of relaxation through phononic channels. This approach, which represents the state-of-the-art of the level of numerical accuracy that can be currently reached, is computationally very  demanding. Also, our understanding is that it allows little flexibility on the choice of the external time-dependent modulation.  

A different approach was pursued in Ref.~\onlinecite{Tsuji15}. Here the external field was taken care of through a time-dependent electronic band dispersion resulting from the direct action of the electromagnetic field on electronic subsystem. Two scenarios were discussed. In the first case (weak-field regime), the interaction was taken to be of the standard BCS form, and the equations of motion for the Anderson pseudospins~\cite{Anderson58} were solved analytically up to second order in the vector potential describing the external field. In the second case, the interaction was taken to be of the Hubbard form and the dynamics of the superconducting order parameter was calculated numerically by using a dynamical mean-field theory approach in a one-dimensional system and in infinite dimensions, assuming a monochromatic oscillating time-dependent field, as in the Floquet formalism.

The purpose of this work is to lay down a nonequilibrium theory of superconductivity that allows us to bypass the aforementioned limitations. More precisely, we neither want to rely on a smallness assumption for the amplitude of the external field nor assume that the external field is periodic in time. The only assumption we want to make is that the field is slowly-varying in time, in the sense discussed above and as will be rigorously formalized below. In this adiabatic regime, we can employ the recently developed Adiabatic Perturbation Theory (APT)~\cite{Rigolin08}. This is a very general procedure that allows to deal with systems whose Hamiltonians have a slow dependence on time, while going systematically beyond the conventional adiabatic theorem of quantum mechanics, which represents the ``zeroth order'' of APT. Applying APT to our nonequilibrium superconducting problem, we are able to lay down a theory which falls into the category (1) of our previous list.

Of course, several theoretical treatments of superconductivity in the adiabatic regime are available. For example, in Ref.~\onlinecite{Kopnin} one can find a microscopic derivation of the time-dependent gap $\Delta(\omega)$ in the frequency representation and in a small-$\omega$ expansion. This derivation is heavily based on the strong assumption that the energy spectrum and quasiparticle distribution function remain the same as at equilibrium. More accurately, the phenomenological Ginzburg-Landau theory, which is applicable at equilibrium for temperatures $\approx T_{\rm c}$, can be extended to nonequilibrium systems, yielding the time-dependent Ginzburg-Landau (TDGL) theory~\cite{Abrahams66, Schmid66, Gorkov68}. The latter is designed to describe systems with temperature close to $T_{\rm c}$ and subject to {\it small} deviations from equilibrium. Approximate differential equations for the time-dependent gap are obtained from the general ones upon expanding in a Taylor series the time and space variations of $\Delta$ with respect to the equilibrium value~\cite{Abrahams66}. As such, this framework cannot describe large variations of the gap parameter.

Our main results are summarized into two equations, which determine the leading contributions to the nonequilibrium gap parameter within the framework of APT [Eqs.~\eqref{Equation Delta0} and~\eqref{Equation Delta1}]. These can be easily solved by elementary numerical approaches. Such equations require the external field to be slowly varying in time (as specified above), but are neither restricted to small variations of the gap nor to periodic external drivings. The derivation will be reported in great detail and can be summarized as follows. We start by describing our system by means of a Hamiltonian that includes electrons and phonons, 
a Fr\"ohlich-type electron-phonon interaction, and a time-dependent external field acting on the phonon subsystem (Section~\ref{sec: Hamiltonian}). We then apply the nonequilibrium path-integral formalism on the Kostantinov-Perel' (KP) time contour~\cite{SecchiPolini1} to derive an effective electronic action, which is obtained after integrating out the phononic degrees of freedom (Section~\ref{sec: noneq path integrals}). The equilibrium version of this approach is standard for stationary superconductivity~\cite{Moulopoulos99, Gersch08}. Several notable differences, however, appear in the nonequilibrium case. In particular, one directly obtains the effective retarded electron-electron interaction in the real-time representation~\cite{Mahan} (Section~\ref{subsec: effective elel}) and, obviously, an action term accounting for the external field (Section~\ref{subsec: external}). We then introduce {\it sources} that enable the calculation of the Cooper parameters by functional differentiation (Section~\ref{subsec: sources}). At this stage, we proceed by approximating the effective electron-phonon interaction {\it \`{a} la} Hubbard (Section~\ref{sec: BCS}). This allows us to perform a Hubbard-Stratonovich decoupling and to integrate out the fermionic degrees of freedom (Section~\ref{subsec: HS}). This procedure yields path-integral expressions for the time-dependent Cooper parameters and, in principle, other observables (Section~\ref{subsec: path integral Cooper}). In practice, these expressions should be evaluated under the nonequilibrium saddle-point approximation (Section~\ref{subsec: SP}), which yields the nonequilibrium version of Gor'kov equations~\cite{Gorkov58}. We finally proceed to determine the self-consistent equation for the nonequilibrium superconducting gap, in the case of a spatially-uniform external field (Section~\ref{sec: spatially uniform}). It is exactly at this step that we utilize APT. The main results of this work are presented in Section~\ref{subsec: Results}, in Eqs.~\eqref{Equation Delta0} and~\eqref{Equation Delta1}, in a form that is easily tractable numerically. In Section~\ref{subsec: recover equilibrium} we show that, at equilibrium, our nonequilibrium formulas reduce to the BCS result. In Section~\ref{sec: necessity and validity} we discuss why the APT approach was necessary and how to assess its validity. In Section \ref{sec: example} we derive the analytical, closed-form solution of Eqs.~\eqref{Equation Delta0} and \eqref{Equation Delta1} at zero initial temperature, and we report a summary of our main numerical results. A summary and a set of conclusions is reported in Section~\ref{sec: conclusions}. A number of relevant technical details can be found in Appendices~\ref{app: bosonic int}-\ref{app: simplifications Delta_1}.

\section{Hamiltonian of the coupled electron-phonon system}
\label{sec: Hamiltonian}

\subsection{Electronic representation}

We consider a system of electrons and phonons described by the following time-dependent Hamiltonian:
\begin{align}\label{Hamiltonian}
\hat{\cal H}(t) & \equiv \sum_{\boldsymbol{k} , \sigma} [\epsilon^{(0)}_{\boldsymbol{k}, \sigma} - \mu_{\sigma}] \hat{c}^{\dagger}_{\boldsymbol{k}, \sigma} \hat{c}_{\boldsymbol{k}, \sigma}       + \sum_{\boldsymbol{q} , \lambda} \omega_{\boldsymbol{q} , \lambda} \hat{b}^{\dagger}_{\boldsymbol{q} , \lambda} \hat{b}_{\boldsymbol{q} , \lambda} \nonumber \\
& \quad  +  \hat{\cal H}_{\mathrm{ep}}   + \hat{\cal H}_{\mathrm{ext}}(t)~ . 
\end{align}
In Eq.~\eqref{Hamiltonian}, the first term is the free-electron Hamiltonian, where $\sigma = \uparrow,\downarrow \equiv \pm 1$, $\epsilon^{(0)}_{\boldsymbol{k} , \sigma}$ is the single-electron energy dispersion, and $\mu_{\sigma}$ is the (possibly spin-dependent) chemical potential. The second term is the free-phonon Hamiltonian, where $\lambda$ labels the phonon branches. The third term,
\begin{align}
\hat{\cal H}_{\mathrm{ep}} \equiv  \sum_{\boldsymbol{q} , \lambda}  M_{\boldsymbol{q}  , \lambda} ( \hat{b}_{\boldsymbol{q} , \lambda}     + \hat{b}^{\dagger}_{- \boldsymbol{q}, \lambda}) \sum_{\boldsymbol{k} , \sigma} \hat{c}^{\dagger}_{\boldsymbol{q} + \boldsymbol{k}, \sigma} \hat{c}_{\boldsymbol{k} , \sigma}~,
\label{el-ph Hamiltonian}
\end{align}
is the electron-phonon interaction Hamiltonian~\cite{AltlandSimons}. Finally, the fourth term,  
\begin{align}
\hat{\cal H}_{\mathrm{ext}}(t)        \equiv \sum_{\boldsymbol{q} , \lambda}  F_{\boldsymbol{q} , \lambda}(t)  ( \hat{b}_{\boldsymbol{q} , \lambda}     + \hat{b}^{\dagger}_{- \boldsymbol{q}, \lambda})~, 
\label{external field}
\end{align}
describes a time-dependent external field displacing the ions from their equilibrium positions. For $\hat{\cal H}(t)$ to be Hermitian, it must be
\begin{align}
M_{\boldsymbol{q} , \lambda}    =  M^*_{- \boldsymbol{q} , \lambda} ~ , \quad \quad F_{\boldsymbol{q} , \lambda}(t) =  F^*_{- \boldsymbol{q} , \lambda}(t)  ~ .
\label{hermiticity}
\end{align}

A mechanism that generates $\hat{\cal H}_{\mathrm{ext}}(t)$ in the form of Eq.~\eqref{external field} could be a nonlinear coupling between infrared-active (IRA) and Raman-active (RA) phonons~\cite{Subedi14, Knap16, Nicoletti16, Sentef16}, the latter being responsible for conventional superconductivity via their interaction with the conduction electrons, while IRA phonons at zero momentum can be coherently excited by a laser. 

Several types of nonlinearities have been recently discussed in great detail in Refs.~\onlinecite{Knap16} and \onlinecite{Nicoletti16}. For example, the phonon Hamiltonians responsible for so-called type-I and type-II nonlinearities can be written, in first quantization, as
\begin{align}\label{I-nonlinearity term}
\hat{\cal H}_{\rm I} = \Lambda_{\rm I} \left( Q^{\mathrm{IRA}}_{\boldsymbol{0}} \right)^2 Q^{\mathrm{RA}}_{\boldsymbol{0}}
\end{align}
and
\begin{align}\label{II-nonlinearity term}
\hat{\cal H}_{\rm II} = \sum_{\boldsymbol{k}} \Lambda_{{\rm II}, \boldsymbol{k}} \left( Q^{\mathrm{IRA}}_{\boldsymbol{0}} \right)^2 Q^{\mathrm{RA}}_{\boldsymbol{k}} Q^{\mathrm{RA}}_{- \boldsymbol{k}}~,
\end{align}
respectively, where $Q^{\mathrm{IRA (RA)}}_{\boldsymbol{q}}$ is the IRA (RA) phonon displacement operator at wave vector $\boldsymbol{q}$ (we have neglected the band index for simplicity). Ref.~\onlinecite{Knap16} mostly focuses on type-II nonlinearities. Here, instead, we concentrate on a type-I phonon nonlinearity. If the IRA phonon field is treated classically and driven coherently by an external electromagnetic field~\cite{Subedi14}, while the RA phonon field is treated quantum-mechanically, i.e.~ $Q^{\mathrm{RA}}_{\boldsymbol{0}} \propto ( \hat{b}_{\boldsymbol{0} , \mathrm{RA}}     + \hat{b}^{\dagger}_{ \boldsymbol{0}, \mathrm{RA}})$, Eq.~\eqref{I-nonlinearity term} coincides with the $\boldsymbol{q} = \boldsymbol{0}$ term in Eq.~\eqref{external field}. Later in our derivation (Section~\ref{sec: spatially uniform}), we will take phonon modes at $\boldsymbol{q} = \boldsymbol{0}$ (which is justified by the smallness of the photon momentum with respect to the reciprocal-lattice vector~\cite{Knap16}), although we develop the first part of the theory in full generality. 

To establish a relationship with previous works, we note the following. As shown in Appendix A of Ref.~\onlinecite{Knap16}, if the feedback of the electrons on the phonon subsystem is neglected, one can treat the nonlinear term given in Eq.~\eqref{I-nonlinearity term} classically, i.e.~by replacing $Q^{\mathrm{RA}}_{\boldsymbol{0}} \rightarrow Q^{\mathrm{RA}}_{\boldsymbol{0}}(t)$ where $Q^{\mathrm{RA}}_{\boldsymbol{0}}(t)$ is determined by a pumped oscillator equation of motion (EOM), pumping being provided by the coherently excited IRA mode. If the analytical solution of this EOM is  inserted in our Eq.~\eqref{el-ph Hamiltonian} in place of the second-quantized phonon operators, one obtains a replacement of $\hat{\cal H}_{\mathrm{ep}}$ with an effective time-dependent single-electron Hamiltonian. Leaving aside the specific choice of the time dependence of this term, this approach is equivalent to that of Ref.~\onlinecite{Sentef16}. Here, however, we treat the RA phonons as quantum fields (see Eq.~\eqref{el-ph Hamiltonian}), and we do not fix {\it a priori} the time dependence of the external field. 

\subsection{Nambu representation}

We now apply the Nambu transformation on the fermionic fields~\cite{Gersch08},
\begin{align}\label{Nambu}
&  \hat{c}_{\boldsymbol{k}, \uparrow} \equiv \hat{d}_{\boldsymbol{k}, \uparrow} ~, \quad  \hat{c}^{\dagger}_{\boldsymbol{k}, \uparrow} \equiv \hat{d}^{\dagger}_{\boldsymbol{k} , \uparrow} ~ , \nonumber \\
&  \hat{c}_{\boldsymbol{k}, \downarrow} \equiv \hat{d}^{\dagger}_{- \boldsymbol{k}, \downarrow} ~, \quad  \hat{c}^{\dagger}_{\boldsymbol{k}, \downarrow} \equiv  \hat{d}_{- \boldsymbol{k}, \downarrow}~,
\end{align}
and we re-define the boson fields as~\cite{Mahan}
\begin{align}
\hat{b}_{\boldsymbol{q} , \lambda} \equiv \hat{a}_{\boldsymbol{q} , \lambda} - \delta_{\boldsymbol{q}, \boldsymbol{0}}   \mathcal{N}   M_{\boldsymbol{0}, \lambda} \, / \, \omega_{\boldsymbol{0}, \lambda} ~,
\end{align}
where $\mathcal{N}$ is the number of $\boldsymbol{k}$ points in the first Brillouin zone. After this substitution, the Hamiltonian in Eq.~\eqref{Hamiltonian} becomes
\begin{align}\label{Hamiltonian Nambu}
\hat{\cal H}(t) & = \sum_{\boldsymbol{k}, \sigma} \sigma \epsilon_{\sigma \boldsymbol{k}, \sigma} \hat{d}^{\dagger}_{\boldsymbol{k} , \sigma} \hat{d}_{\boldsymbol{k} , \sigma} 
     +  \sum_{\boldsymbol{q} ,  \lambda} \omega_{\boldsymbol{q} , \lambda}   \hat{a}^{\dagger}_{\boldsymbol{q} , \lambda}      \hat{a}_{\boldsymbol{q} , \lambda}  \nonumber \\
& \quad + \sum_{\boldsymbol{q} , \lambda} [ M_{\boldsymbol{q} ,  \lambda}  \hat{\rho}_{\boldsymbol{q}} +  F_{\boldsymbol{q}  , \lambda}(t)    ] ( \hat{a}_{\boldsymbol{q} , \lambda}   + \hat{a}^{\dagger}_{-\boldsymbol{q} , \lambda})~.
\end{align}
In writing Eq.~\eqref{Hamiltonian Nambu}, we have: i) introduced
\begin{align}
   \hat{\rho}_{\boldsymbol{q}}  =    \sum_{\boldsymbol{k} , \sigma } \sigma  \hat{d}^{\dagger}_{\boldsymbol{q} + \boldsymbol{k}, \sigma} \hat{d}_{\boldsymbol{k} , \sigma}~, 
\label{rho_q Nambu expression}
\end{align}
ii) defined the renormalized single-electron band energies by
\begin{equation}
\epsilon_{\sigma \boldsymbol{k}, \sigma} \equiv \epsilon^{(0)}_{\sigma \boldsymbol{k}, \sigma}  - \mu_{\sigma}  - 2 \mathcal{N}  \sum_{  \lambda}          (    M^2_{\boldsymbol{0}, \lambda} \, / \, \omega_{\boldsymbol{0}, \lambda})~, 
\label{effective Nambu band}
\end{equation} 
and iii) discarded a time-dependent quantity which involves no operators and, hence, can be gauged away via a common time-dependent phase factor for all wave functions, giving no contribution to the calculations of observables.

\section{Nonequilibrium superconductivity in the path-integral formalism}
\label{sec: noneq path integrals}
\subsection{Partition function and action}
Rather than solving numerically~\cite{Sentef16} the EOMs for the nonequilibrium Green's functions (GFs) for a chosen time-dependent external field, we here develop a semi-analytical approach that allows us to derive an easily-solvable equation for the time-dependent gap parameter. 

In order to do so, we need to make some simplifying assumptions, without loosing certain essential nonequilibrium features. We use the nonequilibrium path-integral formalism on the KP time contour, which enables us to choose initial states of arbitrary nature, to integrate away the phononic degrees of freedom. While nonequilibrium path integrals on the Schwinger-Keldysh contour are thoroughly discussed in Ref.~\onlinecite{Kamenev}, their version on the KP time contour has not been studied with the same level of rigor. All necessary technical details can, however, be found in Ref.~\onlinecite{SecchiPolini1}, whose formalism is employed also in this work. We take $\hbar = 1$ throughout this Article.

At the initial time $t = t_{0}$ the system is described by a known state or statistical mixture, specified by the inverse temperature $\beta$ and the density matrix operator $\hat{n}_{0}(\beta)$. The physical time domain is $t \in [t_{0},  \infty)$. The KP time contour $\gamma$ is then given by the union of three branches: $\gamma = \gamma_+ \cup \gamma_- \cup \gamma_{\rm M}$. The forward ($\gamma_+$) and backward ($\gamma_-$) branches result from doubling the real time degrees of freedom along $[ t_0,  \infty)$. For a given physical time value $t$, we denote by the symbols $t_+$ and $t_-$ the two corresponding contour variables on $\gamma_+$ and $\gamma_-$, respectively. The initial density matrix is written as $\hat{n}_{0}(\beta) = \hat{\cal U}_{\gamma_{\rm M}} / \mathrm{Tr} ( \hat{\cal U}_{\gamma_{\rm M}})$, where $\hat{\cal U}_{\gamma_{\rm M}}$ is the evolution operator along the imaginary-time (Matsubara) branch $\gamma_{\rm M} = [t_0, t_0 - i \beta)$. In the path-integral formalism, the nonequilibrium partition function is written as   
\begin{align}
  Z\left[ V \right]   \equiv \frac{1}{\mathrm{Tr}\left( \hat{\cal U}_{\gamma_{\rm M}} \right)} \int \mathcal{D}(\overline{d}, d) \int \mathcal{D}( a^*, a) 
\mathrm{e}^{i  S[V; \overline{d}, d ; a^*, a]}~, 
\label{Z KB}
\end{align}
which is a functional of a fermionic {\it source} potential $\hat{V}(z)$, which depends on the contour variable $z$. If $\hat{V}(t_{+}) = \hat{V}(t_{-})$, then $Z = 1$. The functional integration runs over the Grassmann numbers $d_{\mathbf{k}, \sigma}(z)$, $\overline{d}_{\mathbf{k}, \sigma}(z)$, and the complex numbers $a_{\mathbf{q}, \lambda}(z)$, $a^*_{\mathbf{q}, \lambda}(z)$, corresponding to the fermionic and the bosonic operators of the system (in the Nambu representation), respectively.

The nonequilibrium action $S[V; \overline{d}, d ; a^*, a]$ is a functional of the source potential, as well as of the field variables (in the following, we will not denote the latter dependence explicitly). For the Hamiltonian \eqref{Hamiltonian Nambu}, the action is given by
\begin{align}
S[V] \equiv S_{\mathrm{e}}[V] +   S_{\mathrm{ep}} ~,
\label{action KB}
\end{align}
where
\begin{align}
S_{\mathrm{e}}[V] = & \iint_{\gamma} \mathrm{d}z \mathrm{d}z' \sum_{\boldsymbol{k}, \sigma} \overline{d}_{\boldsymbol{k}, \sigma}(z) \, \hat{G}^{\mathrm{fe} \, -1}_{\boldsymbol{k}, \sigma}(z, z')  \, d_{\boldsymbol{k}, \sigma}(z') \nonumber \\
& -  \int_{\gamma} \mathrm{d} z   V[\overline{d}(z), d(z) ; z]  
\label{S_e^V}
\end{align}
involves only electronic fields, while $S_{\mathrm{ep}}$ involves the phonon fields and their coupling to the electronic fields, i.e.,
\begin{align}
S_{\mathrm{ep}}   \equiv \sum_{\boldsymbol{q} , \lambda}   S_{\mathrm{ep}; \boldsymbol{q} , \lambda}~,
\end{align}
with
\begin{align}
  S_{\mathrm{ep}; \boldsymbol{q} , \lambda} 
& =  \iint_{\gamma} \mathrm{d}z \mathrm{d}z'    a^*_{\boldsymbol{q}, \lambda}(z) \,  \hat{G}^{\mathrm{fp} \, -1}_{\boldsymbol{q}, \lambda}(z, z')  \, a_{\boldsymbol{q}, \lambda}(z')   \nonumber \\
  & \quad - \int_{\gamma} \mathrm{d} z        \left[ M_{\boldsymbol{q} , \lambda}   \rho_{\boldsymbol{q}}(z)    + F_{\boldsymbol{q} , \lambda}(z)   \right]     a_{\boldsymbol{q} , \lambda}(z)         \nonumber \\
&  \quad - \int_{\gamma} \mathrm{d} z     \left[ M_{- \boldsymbol{q} ,  \lambda}   \rho_{- \boldsymbol{q}}(z)  + \! F_{- \boldsymbol{q} , \lambda}(z)   \right]      a^{*}_{  \boldsymbol{q} , \lambda}(z)      ~. 
\label{S_ph}
\end{align}
The operators $\hat{G}^{\mathrm{fe} \, -1}_{\boldsymbol{q}, \lambda}(z, z')$ and $\hat{G}^{\mathrm{fp} \, -1}_{\boldsymbol{q}, \lambda}(z, z')$ appearing in Eqs.~\eqref{S_e^V} and~\eqref{S_ph} are the {\it{inverse}} free-electron (fe) and free-phonon (fp) GFs, respectively, defined on the contour $\gamma$. Their features are discussed in full generality in Ref.~\onlinecite{SecchiPolini1}. In the case at hand, they are diagonal in the single-particle quantum labels. In Eq.~\eqref{S_ph}, the symbol $\rho_{\boldsymbol{q}}(z)$ denotes the Grassmann representation of the density operator given by Eq.~\eqref{rho_q Nambu expression}, i.e.
\begin{align}
    \rho_{\boldsymbol{q}}(z)  =    \sum_{\boldsymbol{k} , \sigma } \sigma  \overline{d}_{\boldsymbol{q} + \boldsymbol{k}, \sigma}(z) \, d_{\boldsymbol{k} , \sigma}(z)~.
\label{rho Grassmann}
\end{align}
\subsection{Effective electronic action}

Since the action \eqref{S_ph} is quadratic in the phonon fields, we can integrate them away. The Gaussian functional integral is carried out in Appendix~\ref{app: bosonic int}. After the integration, Eq.~\eqref{Z KB} reduces to 
\begin{align}\label{partition function only electrons}
Z[V] = \frac{1}{\mathrm{Tr}\left( \hat{\cal U}_{\gamma_{\rm M}}^{\mathrm{eff}} \right) }  \int \mathcal{D}(\overline{d}, d)
\mathrm{e}^{i S_{\mathrm{eff}}[V]}~,
\end{align}
where we have introduced the following {\it effective electronic action}:
\begin{align}
& S_{\mathrm{eff}}[V] =  S_{\mathrm{e}}[V] + S_{\mathrm{int}}  + S_{\mathrm{ext}} ~.
\label{effective action}
\end{align}
This consists of three terms: the electronic action $S_{\mathrm{e}}[V]$, given by Eq.~\eqref{S_e^V}, and two terms coming from the bosonic integration (as detailed in Appendix~\ref{app: bosonic int}), i.e.~the effective electron-electron interaction $S_{\mathrm{int}}$ and the phonon-mediated coupling between electrons and the external field $S_{\mathrm{ext}}$. The last two contributions are expressed in terms of the {\it{direct}} free-phonon GF, which inverts the operator $\hat{G}^{\mathrm{fp} \, -1}_{\boldsymbol{q}, \lambda}(z, z')$ on the $\gamma$ contour, and is given by~\cite{SecchiPolini1} 
\begin{align}
 G^{\mathrm{fp}}_{\boldsymbol{q}, \lambda}(z, z') = - i \, \mathrm{e}^{- i \, \omega_{\boldsymbol{q}, \lambda} \left(t - t' \right) } \left[ \Theta\left(z, z' \right) + n^{(\rm B)}_{\boldsymbol{q}, \lambda}  \right]~.
\label{Gfp on the contour}
\end{align}
Here, $t$ and $t'$ are complex time coordinates corresponding to the contour coordinates $z$ and $z'$, respectively, $\Theta(z,z')$ is the step function on $\gamma$, with $\Theta(z,z) = 1$, and
\begin{align}
 n^{({\rm B})}_{\boldsymbol{q}, \lambda} \equiv \left( \mathrm{e}^{  \beta \omega_{\boldsymbol{q}, \lambda}}  -  1 \right)^{-1}           
\end{align}
is the bosonic occupation number.

The two phonon-mediated contributions to the effective action in Eq.~\eqref{effective action} are
\begin{align}\label{S eff int}
S_{\mathrm{int}}   &  =    -   \sum_{\boldsymbol{q}, \lambda}   \left| M_{ \boldsymbol{q}, \lambda } \right|^2   \iint_{\gamma}  \mathrm{d} z   \mathrm{d} z'      \rho_{ \boldsymbol{q}}(z)    G^{\mathrm{fp}}_{ \boldsymbol{q}, \lambda}(z, z')  \rho_{- \boldsymbol{q}}(z')
\end{align}
and 
\begin{align}
S_{\mathrm{ext}}     \equiv -  \sum_{\boldsymbol{q} }  \int_{\gamma} \mathrm{d} z     f_{\boldsymbol{q}}(z)  \rho_{  \boldsymbol{q}}(z)~ ,
 \label{S eff ext}
\end{align}
where
\begin{align}
f_{\boldsymbol{q}}(z)  &  \equiv     \sum_{\lambda} M_{  \boldsymbol{q}, \lambda }    \int_{\gamma}   \mathrm{d} z'   [       G^{\mathrm{fp}}_{ \boldsymbol{q}, \lambda}(z, z')     
      +       G^{\mathrm{fp}}_{ - \boldsymbol{q}, \lambda}(z',   z)]  \nonumber \\
      & \quad \times   F_{- \boldsymbol{q}, \lambda}(z')~.  
      \label{f_q(z)}
\end{align}
In Eq.~\eqref{partition function only electrons}, we have also introduced the effective evolution operator $\hat{\cal U}_{\gamma_{\rm M}}^{\mathrm{eff}}$ along $\gamma_{\rm M}$ for the electrons only, which originates from the bosonic integration in the path-integral representation of the quantity $\mathrm{Tr}( \hat{\cal U}_{\gamma_{\rm M}})$. Details are given in Appendix~\ref{app: bosonic int}. In our main derivation, we will not need its explicit expression. It is enough to recall that it is a constant that ensures the normalization $Z[V = 0] = 1$.

In the next two Sections we discuss the additional terms of the effective electronic action.

\subsection{Effective electron-electron interactions}
\label{subsec: effective elel}

The effective action given in Eq.~\eqref{S eff int} describes a phonon-mediated interaction between electrons. To show the correspondence with the well-known BCS retarded interaction, we use Eq.~\eqref{Gfp on the contour} and transform $\rho_{ \boldsymbol{q}}(z)$ (as given in Eq.~\eqref{rho Grassmann}, specified to the cases $z = t_+$, $z = t_-$, and $z = t_0 - i \tau$) via the Keldysh rotation, i.e. 
\begin{align}
   \rho_{ \boldsymbol{q}}(t_{\pm})     \equiv   2^{-1/2} \left[ \rho^{\rm C}_{ \boldsymbol{q}}(t) \pm \rho^{\rm Q}_{ \boldsymbol{q}}(t)   \right]  , \quad   \rho_{ \boldsymbol{q}}(t_0 - i \tau) \equiv \rho^{\rm M}_{ \boldsymbol{q}}(\tau) ~.
   \label{Keldysh rotation}
\end{align}
Eq.~\eqref{Keldysh rotation} defines the classical [$\rho^{\rm C}_{ \boldsymbol{q}}(t)$], quantum [$\rho^{\rm Q}_{ \boldsymbol{q}}(t)$], and Matsubara [$\rho^{\rm M}_{ \boldsymbol{q}}(\tau)$] components of the electronic density operator expressed in Grassmann variables. We obtain
\begin{widetext}
\begin{align}
  S_{  \mathrm{int}}  
&  =  -2  \sum_{\boldsymbol{q}, \lambda} \left| M_{ \boldsymbol{q}, \lambda } \right|^2    \iint_{t_0}^{\infty} \mathrm{d} t   \mathrm{d} t'  \Theta(t' - t)  \,   \sin\left[ \omega_{\boldsymbol{q}, \lambda} \left(t - t' \right) \right]  \rho^{\rm C}_{ \boldsymbol{q}}(t)     \, \rho^{\rm Q}_{- \boldsymbol{q}}(t')    \nonumber \\
& \quad +  i  \sum_{\boldsymbol{q}, \lambda} \left| M_{ \boldsymbol{q}, \lambda } \right|^2    \left(     2  n^{({\rm B})}_{ \boldsymbol{q}, \lambda} + 1    \right)  \int_{t_0}^{\infty} \mathrm{d} t \mathrm{e}^{- i   \omega_{\boldsymbol{q}, \lambda}  t   }         \rho^{\rm Q}_{ \boldsymbol{q}}(t) \int_{t_0}^{\infty}  \mathrm{d} t'  \mathrm{e}^{  i   \omega_{\boldsymbol{q}, \lambda}   t'   }             
   \rho^{\rm Q}_{- \boldsymbol{q}}(t')    \nonumber \\
& \quad +   \sqrt{2} \sum_{\boldsymbol{q}, \lambda}   \left| M_{ \boldsymbol{q}, \lambda } \right|^2   \int_{0}^{ \beta} \mathrm{d} \tau  \int_{t_0}^{\infty} \mathrm{d} t  \left[  \left( 1 + n^{({\rm B})}_{\boldsymbol{q}, \lambda}  \right)  \mathrm{e}^{- i \omega_{\boldsymbol{q}, \lambda} \left(t_0 - i \tau - t \right) }  +     n^{({\rm B})}_{\boldsymbol{q}, \lambda}    \mathrm{e}^{- i \omega_{\boldsymbol{q}, \lambda} \left(t - t_0 + i \tau \right) }    \right]       \rho^{\rm M}_{ \boldsymbol{q}}(\tau)  \,    \rho^{\rm Q}_{- \boldsymbol{q}}(t) \nonumber \\
& \quad - i  \sum_{\boldsymbol{q}, \lambda}   \left| M_{ \boldsymbol{q}, \lambda } \right|^2  \iint_{0}^{\beta} \mathrm{d} \tau  \mathrm{d} \tau'   \mathrm{e}^{  \omega_{\boldsymbol{q}, \lambda} \left(  \tau' -  \tau \right) }             \left[ \Theta\left( \tau - \tau' \right) + n^{({\rm B})}_{\boldsymbol{q}, \lambda}  \right]   \rho^{\rm M}_{ \boldsymbol{q}}(\tau)  \, \rho^{\rm M}_{- \boldsymbol{q}}(\tau')~.
\label{effective interaction}
\end{align}
\end{widetext}
The effective time-dependent (retarded) interaction between the electrons is given by the coefficient which couples the real-time densitites $\rho^{{\rm C}}_{\boldsymbol{q}}(t)$ and $\rho^{{\rm Q}}_{-\boldsymbol{q}}(t')$, i.e.
\begin{align}
V_{\boldsymbol{q}}(t - t') \equiv  -2  \sum_{\lambda} \left| M_{ \boldsymbol{q}, \lambda } \right|^2     \Theta(t' - t)  \,   \sin\left[ \omega_{\boldsymbol{q}, \lambda} \left(t - t' \right) \right]~.
\end{align}
Its Fourier transform is
\begin{align}
V_{\boldsymbol{q}}(\omega) & = -2  \sum_{\lambda} \left| M_{ \boldsymbol{q}, \lambda } \right|^2   \lim_{\zeta \rightarrow 0^+}  \frac{ \omega_{\boldsymbol{q}, \lambda} }{\left( \omega - i \zeta \right)^2 - \omega_{\boldsymbol{q}, \lambda}^2}~,
\label{effective interaction frequency}
\end{align}
which correctly reproduces the BCS retarded interaction (see e.g.~Ref.~\onlinecite{Mahan}, where $g_{\boldsymbol{q}, \lambda}   \equiv \sqrt{2 \omega_{\boldsymbol{q}, \lambda}} M_{\boldsymbol{q}, \lambda}$). 

\subsection{Effective action for the coupling with the external field}
\label{subsec: external}

The action in Eq.~\eqref{S eff ext} expresses the indirect effect of the external field on the electron subsystem, which is mediated by the phonons. We now discuss the effective field defined in Eq.~\eqref{f_q(z)}. Using Eq.~\eqref{Gfp on the contour} and the properties
\begin{align}
& F_{ \boldsymbol{q}, \lambda}(t_+) = F_{ \boldsymbol{q}, \lambda}(t_-) = F_{ \boldsymbol{q}, \lambda}(t), \quad F_{ \boldsymbol{q}, \lambda}(t - i \tau) = 0~,
\label{property F on the contour}
\end{align}
we obtain
\begin{align}
f_{\boldsymbol{q}}(z) 
  & =  - i   \sum_{\lambda} M_{  \boldsymbol{q}, \lambda }   \sum_{\nu = \pm} \nu \int_{t_0}^{\infty}   \mathrm{d} t' \left\{        \mathrm{e}^{- i \omega_{\boldsymbol{q}, \lambda} \left(t - t' \right) }   \Theta\!\left(z, t'_{\nu} \right)  \right. \nonumber \\ 
& \quad \left.      +   \,    \mathrm{e}^{ i \omega_{\boldsymbol{q}, \lambda} \left(t - t' \right) }   \Theta\left(t'_{\nu}, z \right)     \right\}   F_{- \boldsymbol{q}, \lambda}(t')~.   
      \label{f_q(z) passage}
\end{align}
If $z$ belongs to any of the real-time branches, i.e.~$z = t_+$ or $z = t_-$, we obtain
\begin{align}
& f_{\boldsymbol{q}}(t_+) = f_{\boldsymbol{q}}(t_- ) \equiv f_{\boldsymbol{q}}(t) \nonumber \\
& = - 2  \sum_{\lambda} M_{  \boldsymbol{q}, \lambda }  \int_{- \infty}^{t} \mathrm{d} t'         \sin\left[ \omega_{\boldsymbol{q}, \lambda} (t - t') \right]  F_{- \boldsymbol{q}, \lambda}(t')~.
\label{f(t)}
\end{align}
On the other hand, if $z$ belongs to the Matsubara branch,
\begin{align}
f_{\boldsymbol{q}}(t_0 - i \tau)     = 0~.
\end{align}
The action term \eqref{S eff ext} then becomes
\begin{align}
S_{\mathrm{ext}}   =  - \sqrt{2} \sum_{\boldsymbol{q} }  \int_{t_0}^{\infty} \mathrm{d} t     f_{\boldsymbol{q}}(t)  \rho^{\rm Q}_{\boldsymbol{q}}(t)~.
\end{align}
This action has the correct causal structure in the sense that the field $F_{{\bm q}, \lambda}(t')$ has an effect on $\rho^{\rm Q}_{\boldsymbol{q}}(t)$ only if $t > t'$. This {\it retarded} effect on the electron subsytem is qualitatively different from the instantaneous effect that would be obtained by coupling a field directly to the electron  rather than the phonon subsystem.

\subsection{Cooper parameters and fermionic sources}
\label{subsec: sources}

We now focus on the Cooper parameters, which are, in general, matrices in wave vector space:
\begin{align}
& C(t)_{\boldsymbol{k}, \boldsymbol{k}'} \equiv \langle \hat{c}_{ - \boldsymbol{k}, \downarrow}(t) \,  \hat{c}_{ \boldsymbol{k}', \uparrow}(t)  \rangle  =  \langle \hat{d}^{\dagger}_{  \boldsymbol{k}, \downarrow}(t) \,  \hat{d}_{ \boldsymbol{k}', \uparrow}(t)  \rangle~, \nonumber \\
& C^{\dagger}(t)_{\boldsymbol{k}, \boldsymbol{k}'} \equiv \langle \hat{c}^{\dagger}_{ \boldsymbol{k}, \uparrow}(t) \, \hat{c}^{\dagger}_{ - \boldsymbol{k}', \downarrow}(t)     \rangle = \langle \hat{d}^{\dagger}_{ \boldsymbol{k}, \uparrow}(t) \, \hat{d}_{  \boldsymbol{k}', \downarrow}(t)\rangle~.
\end{align}
These quantities can be calculated from the partition function via functional differentiation. To this end, we write the source term in the action~\eqref{S_e^V} as
\begin{align}
&   \int_{\gamma} \mathrm{d}z   V\left[\overline{d}(z), d(z) ; z \right]  \nonumber \\
& =   \int_{\gamma} \mathrm{d}z \sum_{\boldsymbol{k}, \boldsymbol{k}'} \sum_{\sigma = \uparrow, \downarrow}   V^{\sigma, - \sigma}_{\boldsymbol{k}, \boldsymbol{k}'}(z) \,  \overline{d}_{  \boldsymbol{k}, \sigma}(z) \,  d_{ \boldsymbol{k}', - \sigma}(z)~,
\end{align}
and the Cooper parameters are obtained as
\begin{align}
C(t)_{\boldsymbol{k}, \boldsymbol{k}'}  
  = \frac{i}{2} \left. \left\{ \frac{\delta Z[V]}{\delta V^{\downarrow, \uparrow}_{\boldsymbol{k}, \boldsymbol{k}'}(t_+)} 
+ \frac{\delta Z[V]}{\delta V^{\downarrow, \uparrow}_{\boldsymbol{k}, \boldsymbol{k}'}(t_-)} \right\} \right|_{V = 0 } ~ ;
\label{observables_1}
\end{align}
$C^{\dagger}(t)_{\boldsymbol{k}, \boldsymbol{k}'}$ has an analogous expression, except for the replacements $\uparrow \rightarrow \downarrow$ and $\downarrow \rightarrow \uparrow$. Calculations of the Cooper parameters require an explicit expression for $Z$, which we now proceed to derive within the well-known ``Hubbard-BCS'' approximation.

\section{Hubbard-BCS approximation}
\label{sec: BCS}
\subsection{Final effective action}
\label{subsec: HS}
To simplify the partition function, we need to perform the integral over the fermionic fields, which requires to decouple the interaction term by introducing suitable (complex) bosonic fields~\cite{AltlandSimons, Kamenev, Moulopoulos99}. The simplest possibility is to adopt the local approximation~\cite{Kamenev} on the effective electron-electron interaction appearing in Eq.~\eqref{S eff int}, i.e.
\begin{align}
-  \sum_{\lambda}   \left| M_{ \boldsymbol{q}, \lambda } \right|^2      G^{\mathrm{fp}}_{ \boldsymbol{q}, \lambda}(z, z')  \to - \frac{U}{2} \delta(z - z')~,
\label{BCS approximation}
\end{align}
for all values of ${\bm q}$, where $U$ is a Hubbard-like parameter expressing the effective strength of the interaction, which is local in space and acts only between electrons with effective band energies in a range $E_{1} < \epsilon_{\sigma \boldsymbol{k}, \sigma} < E_{2}$. Detailed comments on $E_{1,2}$ are reported below in Sect.~\ref{subsec: rough estimate}. The corresponding action term becomes
\begin{align}
&  S_{  \mathrm{int}} \to S_{U}   =    - U \sum_{\boldsymbol{q}  }       \int_{\gamma} \mathrm{d} z\, \overline{\Phi}_{ \boldsymbol{q}}(z)\,  \Phi_{\boldsymbol{q}}(z)~,
\label{Hubbard action}
\end{align}
where we have introduced the Cooper pair fields
\begin{align} 
& \overline{\Phi}_{ \boldsymbol{q}}(z) \equiv \sum_{\boldsymbol{k}} \overline{d}_{\boldsymbol{k}, \uparrow}(z) \, d_{  \boldsymbol{k} + \boldsymbol{q}, \downarrow}(z), \nonumber \\
 &   \Phi_{ \boldsymbol{q}}(z) \equiv \sum_{\boldsymbol{k}} \overline{d}_{ \boldsymbol{k} + \boldsymbol{q}, \downarrow}(z) \, d_{  \boldsymbol{k}, \uparrow}(z)~. 
 \label{Cooper pair fields} 
\end{align}
We now perform a Hubbard-Stratonovich transformation, introducing auxiliary complex fields $\Delta_{\boldsymbol{q}}(z)$ and $\Delta^*_{\boldsymbol{q}}(z)$, which allow to decouple the fermionic interaction term. We then integrate out the fermionic variables, and we are left with an action involving the auxiliary fields only. We refer the reader to Appendix~\ref{app: HS} for all the details, and give here the resulting partition function after the fermionic integration:
\begin{align}
Z[V] = \frac{c}{\mathrm{Tr} \left( \hat{\cal U}^{\mathrm{eff}}_{\gamma_{\rm M}} \right)}          \int \mathcal{D}\left[\frac{\Delta}{U}, \frac{\Delta^*}{U} \right]
    \, e^{ i S_{\mathrm{BCS}}[V]   }~ ,
\label{introducing nonequilibrium action Delta}
\end{align}
where the constant $c$ includes the integration measure~\cite{NegeleOrland}, and the effective BCS nonequilibrium action is
\begin{align}
i S_{\mathrm{BCS}}[V]    \equiv & \, \mathrm{tr} \left[ \mathrm{ln} \left( - i  \hat{G}^{ -1}[V]     \right) \right]    +   \frac{  i }{ U  }   \int_{\gamma}  \mathrm{d} z    \sum_{\boldsymbol{q}}   \left|  \Delta_{\boldsymbol{q}}(z)    \right|^2~ ,
\label{S_BCS}
\end{align}
where we have introduced the inverse BCS electronic GF on $\gamma$ (in the presence of sources), denoted as $\hat{G}^{-1}[V]$, whose matrix elements are
\begin{widetext}
\begin{align}
   \hat{\mathbf{G}}^{  -1  }_{\boldsymbol{k}, z ; \boldsymbol{k}', z'}[V] \equiv   \left( \begin{matrix} \delta_{\boldsymbol{k}, \boldsymbol{k}'}   \hat{G}^{\mathrm{fe} -1}_{ \boldsymbol{k}, \uparrow }(z,z')   - \delta(z,z') f_{\boldsymbol{k} - \boldsymbol{k}'}(z) 
       & \delta(z,z') \left[ \Delta_{\boldsymbol{k}' - \boldsymbol{k}}(z) -    V^{\uparrow \downarrow}_{\boldsymbol{k}, \boldsymbol{k}'}(z) \right] \\
     \delta(z,z') \left[ \Delta^*_{\boldsymbol{k} - \boldsymbol{k}'}(z) -     V^{\downarrow \uparrow}_{\boldsymbol{k}, \boldsymbol{k}'}(z) \right] &     \delta_{\boldsymbol{k}, \boldsymbol{k}'}    \hat{G}^{\mathrm{fe} -1}_{ \boldsymbol{k}, \downarrow }(z , z')  +  \delta(z,z') f_{\boldsymbol{k} - \boldsymbol{k}'}(z) \end{matrix} \right)~.
           \label{G_BCS-1}
\end{align}
\end{widetext}
This is a matrix in the spaces of wave vectors, spins, and contour coordinates. The matrix on the right-hand side of Eq.~\eqref{G_BCS-1} is written explicitly in spin space. In what follows, its individual elements will be denoted by $ \hat{ G }^{  -1  }_{\boldsymbol{k}, \sigma, z ; \boldsymbol{k}', \sigma', z'}[V]$. For the sake of simplicity, we will use the shorthand $ \hat{ G }^{  -1  }_{\boldsymbol{k}, \sigma, z ; \boldsymbol{k}', \sigma', z'}[0] \equiv   \hat{ G }^{  -1  }_{\boldsymbol{k}, \sigma, z ; \boldsymbol{k}', \sigma', z'}  $. The Dirac deltas in Eq.~\eqref{G_BCS-1} should be interpreted as $\delta(z, z')  \rightarrow \delta(z, z' + 0)$---see Appendix \ref{app: HS}.

Using functional differentiation, we now calculate: 1) the Cooper parameters in Eq.~\eqref{observables_1} and 2) the values of the nonequilibrium gap parameters at the saddle point of the action. A general note about functional derivatives in this formalism is reported in Appendix~\ref{app: functional derivatives}.

\subsection{Path-integral expressions for the Cooper parameters}
\label{subsec: path integral Cooper}

We now derive explicit expressions for the nonequilibrium Cooper parameters. Applying Eq.~\eqref{observables_1} to the partition function in the form of Eq.~\eqref{introducing nonequilibrium action Delta}, we find: 
\begin{align}
 C(t)_{\boldsymbol{k}, \boldsymbol{k}'}  & =  - \frac{i}{2}  \frac{c}{\mathrm{Tr} \left( \hat{\cal U}^{\mathrm{eff}}_{\gamma_{\rm M}} \right)}        \int \mathcal{D}\left[\frac{\Delta}{U}, \frac{\Delta^*}{U} \right]
    \, e^{ i S_{\mathrm{BCS} }[0]   } \nonumber \\
    & \quad \times \left\{    G_{\boldsymbol{k}', \uparrow, t_+ ; \, \boldsymbol{k}   , \downarrow , (t+0)_+ } 
      +   G_{\boldsymbol{k}', \uparrow, t_- ; \, \boldsymbol{k} , \downarrow , (t-0)_- }  \right\} ~,
    \label{observables path integral formula 1}
\end{align}
where $t \pm 0$ denotes an instant of time infinitesimally after/before $t$. Therefore, the coordinate $(t + 0)_+$ is reached infinitesimally later than $t_+$ while walking on the forward branch of the contour, and $(t - 0)_-$ is reached infinitesimally later than $t_-$ while walking on the backward branch. Eq.~\eqref{observables path integral formula 1} is exact within the Hubbard-BCS approximation, but it cannot be evaluated without resorting to further approximations.

\subsection{Nonequilibrium saddle point}
\label{subsec: SP}

In order to compute Eq.~\eqref{observables path integral formula 1} and its Hermitian conjugate, we use the saddle-point (SP) approximation, in full analogy to what is done in the framework of non-equilibrium many-body theory on the Schwinger-Keldysh contour~\cite{Kamenev} and standard field-theoretical procedures at equilibrium~\cite{Moulopoulos99}. 

The nonequilibrium saddle points of the action are obtained by finding extrema of Eq.~\eqref{S_BCS} with respect to variations in the fields $\Delta_{\boldsymbol{q}}(z)$ and $\Delta^*_{\boldsymbol{q}}(z)$. Recalling that Eq.~\eqref{observables path integral formula 1} requires $V = 0$, we write down the SP equations:
\begin{align}
  \frac{ \delta \, i S_{\mathrm{BCS}}[0]  }{\delta \Delta^*_{\boldsymbol{q}}(z)}        \stackrel{\mathrm{SP}}{=} 0  ~,  \quad \quad   \frac{ \delta \, i S_{\mathrm{BCS}}[0]  }{\delta \Delta_{\boldsymbol{q}}(z)}       \stackrel{\mathrm{SP}}{=} 0~.
\end{align}
Performing the functional derivatives (see Appendix \ref{app: functional derivatives}) we find
\begin{align}
&    \Delta_{\boldsymbol{q}}( z )        \stackrel{\mathrm{SP}}{=}    i  U    \sum_{\boldsymbol{k}   }            G_{\boldsymbol{k}, \uparrow, z ; \, \boldsymbol{k} + \boldsymbol{q}   , \downarrow, z + 0  }  ~     ,  \nonumber \\
&    \Delta^{*  }_{\boldsymbol{q}}( z )   \stackrel{\mathrm{SP}}{=}   i U  \sum_{\boldsymbol{k}  }      G_{\boldsymbol{k}, \downarrow, z ; \, \boldsymbol{k}  - \boldsymbol{q}   , \uparrow , z + 0 } ~, 
\label{SP equations} 
\end{align}
where $z + 0$ is a contour coordinate occurring infinitesimally later than $z$, for any $z \in \gamma$. In particular, on the Matsubara branch, i.e.~for $z = t_0 - i \tau$, 
\begin{align}
&    \Delta^{({\rm M})}_{\boldsymbol{q}}(t_0 - i \tau)      \stackrel{\mathrm{SP}}{=}    i  U    \sum_{\boldsymbol{k}   }            G_{\boldsymbol{k}, \uparrow, t_0 - i \tau  ; \,  \boldsymbol{k} + \boldsymbol{q}   , \downarrow, t_0 - i (\tau + 0 )  }      ~ ,  \nonumber \\
&   [\Delta^{({\rm M})}_{\boldsymbol{q}}(t_0 - i \tau)]^*   \stackrel{\mathrm{SP}}{=}  i U  \sum_{\boldsymbol{k}  }          G_{\boldsymbol{k}, \downarrow, t_0 - i \tau ; \, \boldsymbol{k}  - \boldsymbol{q}   , \uparrow , t_0 - i (\tau + 0 ) }~.
\label{saddle point M} 
\end{align}

For $z$ belonging to one of the real-time branches ($\gamma_+$ or $\gamma_-$), we introduce the classical (${\rm C}$) and quantum (${\rm Q}$) combinations
\begin{align}
& \Delta^{{\rm C}/{\rm Q}}_{\boldsymbol{q}}(t) \equiv \frac{\Delta_{\boldsymbol{q}}(t_+)   \pm \Delta_{\boldsymbol{q}}(t_-)  }{2} \nonumber \\
& \stackrel{\mathrm{SP}}{=} \frac{ i  U}{2}    \sum_{\boldsymbol{k}   }     \left\{       G_{\boldsymbol{k}, \uparrow, t_+ ; \,  \boldsymbol{k} + \boldsymbol{q}   , \downarrow, (t+0)_+  } 
   \pm     G_{\boldsymbol{k}, \uparrow, t_- ; \,  \boldsymbol{k} + \boldsymbol{q}   , \downarrow , (t-0)_- }  \right\}  , 
\end{align}
\begin{align}
& [\Delta^{{\rm C}/{\rm Q}}_{\boldsymbol{q}}(t)]^{*} \equiv \frac{\Delta^*_{\boldsymbol{q}}(t_+)   \pm \Delta^*_{\boldsymbol{q}}(t_-)  }{2}  \nonumber \\
& \stackrel{\mathrm{SP}}{=}  \frac{ i U }{2} \sum_{\boldsymbol{k}  }   \left\{       G_{\boldsymbol{k}, \downarrow , t_+ ; \, \boldsymbol{k}  - \boldsymbol{q}   , \uparrow , (t+0)_+  } 
\pm   G_{\boldsymbol{k}, \downarrow , t_- ; \, \boldsymbol{k}  - \boldsymbol{q}   , \uparrow , (t-0)_-  }  \right\}    .
\label{saddle point C Q}
\end{align}
From Eq.~\eqref{G_BCS-1} we see that, if $\Delta_{\boldsymbol{q}}(t_+)   = \Delta_{\boldsymbol{q}}(t_-) \equiv \Delta^{(\mathrm{MF})}_{\boldsymbol{q}}(t)$ at all times $t$ and $\Delta^{({\rm M})}_{\boldsymbol{q}}(t_0 - i \tau) \equiv \Delta^{(\rm{MF})}_{\boldsymbol{q}}(t_0)$ for all values of $\tau$, then $G$ becomes a standard nonequilibrium GF corresponding to the following effective mean-field (MF) Hamiltonian:
\begin{widetext}
\begin{align}
\hat{\cal H}_{\rm MF}(t) \equiv \sum_{\boldsymbol{k}, \boldsymbol{k}'} \left( \begin{matrix} \hat{d}^{\dagger}_{\boldsymbol{k}, \uparrow} & \hat{d}^{\dagger}_{\boldsymbol{k}, \downarrow} \end{matrix}\right)   
    \left( \begin{matrix}   \delta_{\boldsymbol{k}, \boldsymbol{k}'}  \epsilon_{\boldsymbol{k}, \uparrow} + f_{\boldsymbol{k} - \boldsymbol{k}'}(t)    &    -   \Delta^{({\rm MF})}_{\boldsymbol{k}' - \boldsymbol{k}}(t)     \\
  -  [\Delta^{({\rm MF})}_{\boldsymbol{k} - \boldsymbol{k}'}(t)]^*    &    -  \delta_{\boldsymbol{k}, \boldsymbol{k}'} \epsilon_{- \boldsymbol{k}, \downarrow} - f_{\boldsymbol{k} - \boldsymbol{k}'}(t)   \end{matrix}  \right)  
  \left( \begin{matrix} \hat{d}_{\boldsymbol{k}', \uparrow} \\ \hat{d}_{\boldsymbol{k}', \downarrow} \end{matrix}\right)~.
  \label{mean-field Hamiltonian}
\end{align} 
\end{widetext}
Here, $\Delta^{({\rm MF})}_{\bm q}(t)$ acts as a classical time-dependent field (since the values of the field are the same, at a given $t$, on both branches of the real-time part of the KP contour). In this situation, one has 
\begin{align}
 G_{t_+ ; \, (t+0)_+} = G_{t_- ; \, (t-0)_-}   \equiv   G^{<  }_{t; \, t}  ~ , 
\end{align}
and we conclude that a self-consistent solution of the SP equations is given by
\begin{align}
& \Delta^{({\rm C})}_{\boldsymbol{q}}(t) \stackrel{\mathrm{SP}}{=} \Delta^{({\rm MF})}_{\bm q}(t)~, \nonumber \\
& \Delta^{({\rm Q})}_{\boldsymbol{q}}(t) \stackrel{\mathrm{SP}}{=} 0~, \nonumber \\
& \Delta^{({\rm M})}_{\boldsymbol{q}}(t_0 - i \tau) \stackrel{\mathrm{SP}}{=} \Delta^{({\rm MF})}_{\bm q}(t_0)~.
\label{Q and M solutions SP }
\end{align}
Eqs.~\eqref{saddle point C Q} should then be solved under the conditions~\eqref{Q and M solutions SP } to determine the MF values $\Delta^{({\rm MF})}_{\bm q}(t)$ corresponding to the saddle points of the BCS action, i.e.
\begin{align}
 \Delta^{({\rm MF})}_{\bm q}(t) & \equiv    i  U  \sum_{\boldsymbol{k}   }          G^{< }_{\boldsymbol{k}, \uparrow, t ; \,  \boldsymbol{k} + \boldsymbol{q}   , \downarrow , t }  ~.
\label{mean-field equation}
\end{align}
The solution of Eq.~\eqref{mean-field equation} automatically satisfies the analogous equation for $\Delta^{({\rm MF}) *}_{\boldsymbol{q}}(t)$~.

In the SP approximation the Cooper-parameter matrix in Eq.~\eqref{observables path integral formula 1} and its Hermitian conjugate are given by
\begin{align}
 & C(t)_{\boldsymbol{k}, \boldsymbol{k}'}   \stackrel{\mathrm{SP}}{=}  -     i        G^{< }_{\boldsymbol{k}', \uparrow , t ; \, \boldsymbol{k}  , \downarrow , t }   ~, \nonumber \\
& C^{\dagger}(t)_{\boldsymbol{k}, \boldsymbol{k}'}   \stackrel{\mathrm{SP}}{=} -      i        G^{<  }_{\boldsymbol{k}', \downarrow, t ; \, \boldsymbol{k} , \uparrow , t  } ~.
    \label{observables SP}
\end{align}
Corrections to Eqs.~\eqref{observables SP} beyond the SP approximation can be included by considering  Gaussian fluctuations of the action around the mean-field point(s). For the application of this procedure to a nonequilibrium problem, we refer the reader to Ref.~\onlinecite{Dunnett16}. This is beyond the scope of the present Article.

\section{The case of a spatially-uniform time-dependent external field}
\label{sec: spatially uniform}

We now consider the case of a uniform time-dependent external field,
\begin{equation}\label{eq:uniform}
f_{\boldsymbol{q}}(t) = \delta_{\boldsymbol{q}, \boldsymbol{0}} f(t)~,
\end{equation}
and look for spatially uniform solutions of the MF equations, i.e.
\begin{equation}
\Delta^{({\rm MF})}_{\bm q}(t) = \delta_{\boldsymbol{q}, \boldsymbol{0}} \Delta(t)~.
\end{equation}
We also assume no magnetic fields acting on the electron subsystem, as well as spatial inversion symmetry, so that $\epsilon_{\sigma \boldsymbol{k}, \sigma} = \epsilon_{\boldsymbol{k}}$ and $\epsilon_{-\bm k} = \epsilon_{\bm k}$~.

In this case, the MF Hamiltonian~\eqref{mean-field Hamiltonian} simplifies to (suppressing in what follows all ``MF'' labels)
\begin{equation}
\hat{\cal H}(t) = \sum_{\boldsymbol{k}} \left( \begin{matrix} \hat{d}^{\dagger}_{\boldsymbol{k}, \uparrow} & \hat{d}^{\dagger}_{\boldsymbol{k}, \downarrow} \end{matrix}\right) \!   
    \left( \begin{matrix}     \epsilon_{\boldsymbol{k}} + f(t)    &    -   \Delta(t)     \\
  -  \Delta^{*  }(t)    &    -  \epsilon_{\boldsymbol{k}} - f(t)   \end{matrix}  \right)  \!
  \left( \begin{matrix} \hat{d}_{\boldsymbol{k}, \uparrow} \\ \hat{d}_{\boldsymbol{k}, \downarrow} \end{matrix}\right)~.
  \label{effective Hamiltonian noneq homogeneous}
\end{equation}
Eq.~\eqref{mean-field equation} becomes
\begin{align}
 \Delta(t)  = - U   \sum_{\boldsymbol{k}} \sum_{\Psi_0}   W_{\Psi_0} \left< \Psi(t) \right| \hat{d}^{\dagger}_{\boldsymbol{k}, \downarrow } \hat{d}_{\boldsymbol{k}, \uparrow } \left| \Psi(t) \right>  ~  ,
\label{saddle point C uniform}
\end{align}
where it is intended that the right-hand side depends functionally on $\Delta(t)$ and $\Delta^*(t)$. The quantities $W_{\Psi_0}$ in the second line of Eq.~(\ref{saddle point C uniform}) are the statistical weights of the states $\left| \Psi_0 \right>$, which are eigenstates of $\hat{\cal H}(t = t_{0})$. The MF states satisfy the time-dependent Schr\"odinger equation,
\begin{align}\label{eq:TDSE}
i \, \partial_t \left| \Psi(t) \right\rangle = \hat{\cal H}(t) \left| \Psi(t) \right\rangle, \quad \left| \Psi(t_0) \right\rangle = \left| \Psi_0 \right\rangle  ~.
\end{align}

The explicit evaluation of $\Delta(t)$ requires an approximate solution of Eq.~\eqref{eq:TDSE}. To make analytical progress, we limit our attention to slowly-varying external fields. This suggests the application of the adiabatic theorem of quantum mechanics~\cite{Messiah}.  However, as detailed in Sect.~\ref{subsec: time der}, the strict application of the adiabatic theorem generates an {\it inconsistency}. On the one hand, it yields a self-consistent equation for $\Delta(t)$. On the other hand, the evaluation of the time derivative $\dot{\Delta}(t)$ via another independent equation yields a vanishing result under the assumptions of the adiabatic theorem. This means that, for the problem at hand, the adiabatic theorem yields a meaningful result only at equilibrium and is therefore useless for our scope.

We now proceed to present a derivation of a time-dependent gap equation in the quasi-adiabatic limit, which bypasses the limitations of the adiabatic theorem.

\subsection{Adiabatic Perturbation Theory}
\label{subsec: APT}

The appropriate tool to deal with systems {\it close} to the adiabatic regime is  {\it Adiabatic Perturbation Theory} (APT)~\cite{Rigolin08}. 
We here summarize the main results of Ref.~\onlinecite{Rigolin08}. 

We rescale the time coordinate $t$ by $T$, which is the time scale over which the system is under observation, introducing the dimensionless quantity $s = t / T$. In the adiabatic regime, the external field (and, therefore, the full Hamiltonian) is assumed to vary slowly on the scale $T$. The time-dependent 
Schr\"odinger equation becomes
\begin{align}\label{eq:rescaled_TDSE}
\frac{i}{T} \partial_s \left| \Psi(s) \right\rangle = \hat{\cal H}(s) \left| \Psi(s) \right\rangle~.
\end{align}
We introduce the instantaneous eigenstates $ \left| n (s) \right>$ of the Hamiltonian:
\begin{align}
& \hat{\cal H}(s) \left| n (s) \right\rangle = \mathcal{E}_{n}(s) \left| n(s) \right\rangle, \nonumber \\
& \left< n(s) |   n'(s) \right> = \delta_{n, n'}~,
\end{align}
for all $s$, where $n$ is the set of quantum numbers specifying $\left | n(s) \right \rangle$. For every $s$, we expand the exact states $\left| \Psi(s) \right\rangle$ solving Eq.~(\ref{eq:rescaled_TDSE}) on the basis of the complete set of states $\left| n(s) \right\rangle$,
\begin{align}
\left| \Psi(s) \right> \equiv \sum_{n} b_{\Psi, n}(s) \mathrm{e}^{ i  \left[ \gamma_n(s) -    \omega_n(s) T \right] } \left| n(s) \right\rangle~ ,
\label{general n vs nI}
\end{align}
where we have introduced the {\it geometrical} phase factor
\begin{align}
\gamma_n(s) = i \int_{s_0}^{s} \mathrm{d} s' \left< n (s') \right| \partial_{s'} \left| n(s') \right\rangle
\label{gamma}
\end{align}
and the {\it dynamical} phase factor
\begin{align}
\omega_n(s) =  \int_{s_0}^s \mathrm{d} s' \mathcal{E}_{n}(s')~.
\label{dynamical phase factor}
\end{align}
The coefficients $b_{\Psi, n}(s)$ are determined by inserting Eq.~\eqref{general n vs nI} into the time-dependent Schr\"odinger equation. We obtain the differential equation
\begin{align}
\partial_s b_{\Psi, n}(s)  =    - \sum_{m \neq n} \mathrm{e}^{i \left[ \gamma_{m, n}(s) - \omega_{m, n}(s) T \right] } M_{n, m}(s) \, b_{\Psi, m}(s)~,
\label{equation b}
\end{align}
where
\begin{align}
&\gamma_{m, n}(s) \equiv \gamma_{m}(s) - \gamma_{n}(s) ~, \nonumber \\
& \omega_{m, n}(s) \equiv \omega_{m}(s) - \omega_{n}(s) ~, \nonumber \\
& M_{n, m}(s) \equiv \left< n(s) \right| \partial_s \left| m(s) \right> = 
\frac{ \left\langle n(s) \right| [ \partial_s \hat{{\cal H}}(s) ] \left| m(s) \right\rangle }{\mathcal{E}_{m }(s) - \mathcal{E}_{n}(s)} \nonumber \\
& \quad \quad \quad \quad = - M_{m, n}^*(s) ~.
\label{adiabatic factor}
\end{align}
The initial condition is
\begin{align}
b_{\Psi, n}(s_0) = \delta_{\Psi_0, n} ~.
\label{initial condition b}
\end{align}
The adiabatic theorem applies exactly if $\partial_s b_{\Psi, n}(s)  = 0$. If the right-hand side of Eq.~\eqref{equation b} is ``small'' (see below), but non-zero, we are in the regime of applicability of APT. This means that the quantities $M_{n, m}(s)$ must be much smaller than unity. This is the requirement of ``slowness'' of the external field mentioned in Sect.~\ref{sec: Intro}. A more rigorous assessment of the validity of APT in our case will be given in Sect.~\ref{subsec: applicability}.

Following Ref.~\onlinecite{Rigolin08}, we make the following Ansatz:
\begin{align}
| \Psi(s)\rangle \equiv \sum_{p = 0}^{\infty} T^{- p}| \Psi^{(p)}(s)\rangle~,
\label{APT ansatz}
\end{align}
where
\begin{align}
| \Psi^{(p)}(s) \rangle \equiv \sum_{n} \mathrm{e}^{i \left[ \gamma_n(s) - \omega_n(s) T\right] } b^{(p)}_{\Psi, n}(s) \left| n(s) \right\rangle
\end{align}
and
\begin{align}
b^{(p)}_{\Psi, n}(s) \equiv \sum_m \mathrm{e}^{- i \left[ \gamma_{n, m}(s) - \omega_{n, m}(s) T\right] } b^{(p)}_{\Psi; n, m}(s)~.
\end{align}
Note that this is equivalent to setting
\begin{align}
b_{\Psi, n}(s) & = \sum_{p = 0}^{\infty} T^{-p} b^{(p)}_{\Psi, n}(s) \nonumber \\
& = \sum_{p = 0}^{\infty} T^{-p}  \sum_m \mathrm{e}^{- i \left[ \gamma_{n, m}(s) - \omega_{n, m}(s) T\right] } b^{(p)}_{\Psi; n, m}(s)~.
\label{b and bb}
\end{align}

After inserting this expansion into the time-dependent Schr\"odinger equation, one obtains equations for the coefficients $b^{(p)}_{\Psi; n, m}(s)$, which can be solved {\it order-by-order}. This is because the equation for $b^{(p)}_{\Psi; n, m}$ at any given $p$ involves only derivatives of these coefficients corresponding to orders $p' < p$. The initial condition is determined from
\begin{align}
\left| \Psi(s_0) \right> = | \Psi^{(0)}(s_0)\rangle~,
\label{initial condition sum}
\end{align}
which follows from Eq.~\eqref{initial condition b}, and by requiring that $| \Psi^{(0)}(s)\rangle$ coincides with the strong adiabatic solution, which is obtained from Eq.~\eqref{equation b} with the right-hand side $=0$. One obtains
\begin{align}
| \Psi^{(0)}(s) \rangle =    \mathrm{e}^{i \left[ \gamma_{\Psi_0}(s) - \omega_{\Psi_0}(s) T\right] } | \Psi_0(s)\rangle 
\label{Psi0 final}
\end{align}
and
\begin{align}
 | \Psi^{(1)}(s) \rangle =  
& \, i  \sum_{n \neq \Psi_0} \left\{ \mathrm{e}^{i \left[ \gamma_{\Psi_0}(s) - \omega_{\Psi_0}(s) T\right] }  \frac{M_{n ,  \Psi_0}(s) }{\mathcal{E}_n(s) - \mathcal{E}_{\Psi_0}(s) } \right. \nonumber \\   
& \left.       -   \mathrm{e}^{i \left[ \gamma_n(s) - \omega_n(s) T\right] }      \frac{M_{n , \Psi_0}(s_0) }{\mathcal{E}_n(s_0) - \mathcal{E}_{\Psi_0}(s_0) }   \right\} \left| n(s) \right> \nonumber \\
&   + i \sum_{  n \neq \Psi_0 } \mathrm{e}^{i \left[ \gamma_{\Psi_0}(s) - \omega_{\Psi_0}(s) T\right] }        J_{n , \Psi_0}(s)  \left| \Psi_0(s) \right>~, 
\label{Psi1 final}
\end{align}
where $\left| \Psi_0(s) \right>$ is the instantaneous eigenstate of $\hat{\cal H}(s)$ that coincides with $\left| \Psi_0 \right>$ at $s = s_0$, and
\begin{align}
 J_{m, n}(s) \equiv \int_{s_0}^s \mathrm{d} s'      \frac{  \left| M_{m, n}(s') \right|^2 }{\mathcal{E}_m(s') - \mathcal{E}_n(s') }~.
\label{Jmn}
\end{align}

We now proceed to calculate Eqs.~\eqref{Psi0 final} and~\eqref{Psi1 final} for our problem. Then, using the resulting expressions, we will derive the leading APT terms of Eq.~\eqref{saddle point C uniform}.

\subsection{Instantaneous eigenstates}
\label{subsect: Instantaneous}

The diagonalization of the uniform mean-field Hamiltonian $\hat{\cal H}(t)$ in Eq.~(\ref{effective Hamiltonian noneq homogeneous}) at each time $t$ yields
\begin{align}
\hat{\cal H}(t) = \sum_{\boldsymbol{k}} \sum_{\alpha = \pm 1} \alpha E_{ \boldsymbol{k}}(t)  \hat{D}^{\dagger}_{\boldsymbol{k}, \alpha}(t) \, \hat{D}_{\boldsymbol{k}, \alpha}(t)~,
\label{eff H diagonalized}
\end{align}
where
\begin{align}
E_{\boldsymbol{k}}(t) \equiv \sqrt{\left[ \epsilon_{\boldsymbol{k}} + f(t) \right]^2 + \left| \Delta(t) \right|^2 } 
\label{instantaneous gapped spectrum}
\end{align}
is the gapped spectrum of the instantaneous quasiparticles (IQPs) corresponding to the time-dependent fermionic fields
\begin{align}
\hat{D}_{\boldsymbol{k}, \alpha}(t) \equiv a^*_{\boldsymbol{k}, \alpha}(t) \hat{d}_{\boldsymbol{k}, \uparrow}  + b^*_{\boldsymbol{k}, \alpha}(t) \hat{d}_{\boldsymbol{k} , \downarrow} , \quad \alpha = \pm 1~.
\label{D fields explicit}
\end{align}
In Eq.~(\ref{D fields explicit})
\begin{align}
&  a_{\boldsymbol{k}, \alpha}(t) =  \frac{- \alpha \Delta(t) }{\sqrt{2 E_{ \boldsymbol{k}}(t)   \left\{ E_{ \boldsymbol{k}}(t) - \alpha \left[ \epsilon_{\boldsymbol{k}} + f(t) \right] \right\}  } }~,  \nonumber \\
&  b_{\boldsymbol{k}, \alpha}(t)  =  \frac{E_{ \boldsymbol{k}}(t) - \alpha \left[ \epsilon_{\boldsymbol{k}} + f(t) \right]  }{\sqrt{2 E_{ \boldsymbol{k}}(t)   \left\{ E_{ \boldsymbol{k}}(t) - \alpha \left[ \epsilon_{\boldsymbol{k}} + f(t) \right] \right\}  } }~.
\label{explicit eigenstates}
\end{align}
At all times, it holds that
\begin{align}
 a^*_{\boldsymbol{k}, \alpha}(t) a_{\boldsymbol{k}, \alpha'}(t) + b^*_{\boldsymbol{k}, \alpha}(t) b_{\boldsymbol{k}, \alpha'}(t) = \delta_{\alpha, \alpha'} ~ ,
 \label{orthonormalization}
\end{align}
and the inverse of Eq.~\eqref{D fields explicit} is
\begin{align}
& \hat{d}_{\boldsymbol{k}, \uparrow} = \sum_{\alpha} a_{\boldsymbol{k}, \alpha}(t) \hat{D}_{\boldsymbol{k}, \alpha}(t) ~, \quad  \hat{d}_{\boldsymbol{k}, \downarrow} = \sum_{\alpha} b_{\boldsymbol{k}, \alpha}(t) \hat{D}_{\boldsymbol{k}, \alpha}(t) ~.
\label{inverse D d}
\end{align}
The instantaneous eigenstates of $\hat{\cal H}(t)$ are then
\begin{align}
\left| n(t) \right> = \prod_{\boldsymbol{k}, \alpha} \left[ \hat{D}^{\dagger}_{\boldsymbol{k}, \alpha}(t) \right]^{n_{\boldsymbol{k}, \alpha} } \left| 0_D \right>~,
\label{instantaneous eigenstates}
\end{align}
where the occupation numbers $n_{\boldsymbol{k}, \alpha} = 1$ or $0$, and $\left| 0_D   \right>$ is the vacuum of all the $\hat{D}$ operators, i.e.
\begin{align}
\hat{D}_{\boldsymbol{k}, \alpha}(t)  \left| 0_D \right\rangle = 0
\end{align}
for all values of $\boldsymbol{k}$, $\alpha$, and $s$. Note that $\left| 0_D \right \rangle$ is also the vacuum of all the $\hat{d}$ operators. As such, it is independent of time. The instantaneous energy eigenvalues are
\begin{align}
\mathcal{E}_{n}(t) = \sum_{\boldsymbol{k}, \alpha} \alpha n_{\boldsymbol{k}, \alpha} E_{ \boldsymbol{k}}(t) ~.
\label{instantaneous spectrum}
\end{align}
In the following, we will put 
\begin{align}
\Delta(s) = \left| \Delta(s) \right| \mathrm{e}^{i \phi(s)}~.
\label{complex gap}
\end{align}
We then have to derive the APT quantities required in Eqs.~\eqref{Psi0 final} and~\eqref{Psi1 final}, specialized to our problem. This involves some lengthy but straightforward algebraic manipulations, whose details are given in Appendix~\ref{app: APT details}.

\subsection{APT expansion of the dynamical gap parameter}

After inserting the APT expansion \eqref{APT ansatz} of the time-dependent states into Eq.~\eqref{saddle point C uniform}, one directly obtains an expansion of $\Delta(s)$ having the form
\begin{align}
 \Delta(s)  \equiv    \sum_{p  = 0}^{\infty} T^{- p } \Gamma_{\Delta}^{(p)}(s)
\label{equation for Delta(s) adiabatic expansion}
\end{align}
with
\begin{align}
\Gamma_{\Delta}^{(p)}(s) \equiv  & \, - U   \sum_{\boldsymbol{k}} \sum_{\Psi_0}   W_{\Psi_0} \nonumber \\
& \times \sum_{q = 0}^p     \langle\Psi^{(q)}(s)  | ~ \hat{d}^{\dagger}_{\boldsymbol{k}, \downarrow } \hat{d}_{\boldsymbol{k}, \uparrow }   ~  | \Psi^{(p - q)}(s)  \rangle~.
\end{align}
We observe that $\Gamma_{\Delta}^{(p)}(s)$ is a functional of $\Delta(s')$ through the dependence of the states $\left| \Psi(s) \right>$ on such quantity. Therefore, $\Gamma_{\Delta}^{(p)}(s)$ itself has a complicated dependence on all powers $T^{-q}$, $q \geq 0$, as follows from Eq.~\eqref{equation for Delta(s) adiabatic expansion}. In turn, this means that truncating the sum in Eq.~\eqref{equation for Delta(s) adiabatic expansion} with respect to $p$ would be incorrect within the framework of APT. What we need is a perturbative expansion of the dynamical gap parameter of the form
\begin{align}
 \Delta(s)  \equiv    \sum_{p  = 0}^{\infty} T^{- p}  \Delta^{(p)}(s)~,
\label{Delta(s) adiabatic expansion}
\end{align}
where the coefficients $\Delta^{(p)}(s)$ do not depend on powers of $T^{-1}$. To determine them, we must 1) explicitly derive a sufficiently large set of quantities $\Gamma_{\Delta}^{(p)}(s)$; 2) insert in these expressions the expansion \eqref{Delta(s) adiabatic expansion}; 3) in Eq.~\eqref{equation for Delta(s) adiabatic expansion}, insert the resulting expressions in the right-hand side, and replace the left-hand side with the expansion \eqref{Delta(s) adiabatic expansion}; 4) identify the terms with the same dependence on $T^{-p}$ on both sides.  

In this Article, we determine the coefficients $\Delta^{(p)}(s)$ corresponding to the lowest values of $p$, i.e.~$p=0$ (strictly adiabatic term) and $p=1$ (first non-adiabatic correction). For this purpose, we only need the quantities $\Gamma_{\Delta}^{(0)}(s)$ and $\Gamma_{\Delta}^{(1)}(s)$, whose derivation is given in Appendix \ref{app: Gamma}. The end result of this procedure is reported in the following Section.

\subsection{Equations for the dynamical gap parameter within first-order APT}
\label{subsec: Results}

We now write down explicit equations for $\Delta^{(0)}(s)$ and $\Delta^{(1)}(s)$. With a minimal abuse of notation, we restore $t = T s$ and replace $\Delta^{(0)}(s) \rightarrow \Delta^{(0)}(t)$ and $T^{-1}\Delta^{(1)}(s) \rightarrow \Delta^{(1)}(t)$. The gap parameter is then obtained as $\Delta(t) \approx \Delta^{(0)}(t) +  \Delta^{(1)}(t)$. From now on, we use the dot to denote $\partial_t$, i.e., $\dot{f}(t) \equiv \partial_t f(t)$. We also attach the subscript $\Delta^{(0)}$ to the quantities which depend functionally on $\Delta^{(0)}(t)$. 

We find that the self-consistent equation for $\Delta^{(0)}(t)$ reads as following:
\begin{align}
\Delta^{(0)}(t)   = -   \Delta^{(0)}(t)   \frac{U}{2}   \sum_{\boldsymbol{k}}   \frac{ w_{\boldsymbol{k}} }{ E_{\Delta^{(0)}, \boldsymbol{k}}(t)   }~,
\label{Equation Delta0}
\end{align}
where
\begin{align}
w_{\boldsymbol{k}} \equiv \sum_{n}   W_{n} \left(   n_{\boldsymbol{k}, -}  -   n_{\boldsymbol{k}, +}    \right)  ~.
\label{w_k}
\end{align}
Once $\Delta^{(0)}(t)$ has been computed, $\Delta^{(1)}(t)$ can be obtained from the following equation:
\begin{widetext}
\begin{align}
  \Delta^{(1)}(t)     & =    \frac{ U  }{4 \left[ 1 + X_{ \Delta^{(0)}}(t) \right] }       \sum_{\boldsymbol{k}}  w_{\boldsymbol{k}}  \left\{  i \, \frac{\Delta^{(0)}(t)  \dot{f}(t) - \left[ \epsilon_{\boldsymbol{k}} + f(t) \right]     \dot{\Delta}^{(0)}(t)   }{  E^3_{\Delta^{(0)}, \boldsymbol{k}}(t) }     \right. \nonumber \\  
          & \quad + \Delta^{(0)}(t) \left[ i \sin\left[ \theta_{\Delta^{(0)}, \boldsymbol{k}}(t) \right]   - \frac{     \epsilon_{\boldsymbol{k}} + f(t)   }{   E_{\Delta^{(0)}, \boldsymbol{k}}(t)     } \cos\left[ \theta_{\Delta^{(0)}, \boldsymbol{k}}(t) \right]  \right]    \frac{ {\rm Im} \! \left[  \dot{\Delta}^{(0)}(t_0) / \Delta^{(0)}(t_0)  \right] }{  E^2_{\Delta^{(0)}, \boldsymbol{k}}(t_0)      }    \nonumber \\
          & \quad \left.   - \Delta^{(0)}(t)   \left[    i  \cos\left[ \theta_{\Delta^{(0)}, \boldsymbol{k}}(t) \right]  + \frac{     \epsilon_{\boldsymbol{k}} + f(t)   }{   E_{\Delta^{(0)}, \boldsymbol{k}}(t)     }  \sin\left[ \theta_{\Delta^{(0)}, \boldsymbol{k}}(t) \right]  \right]  \!  \frac{\dot{f}(t_0) - \left[ \epsilon_{\boldsymbol{k}} + f(t_0) \right] {\rm Re} \! \left[  \dot{\Delta}^{(0)}(t_0) / \Delta^{(0)}(t_0) \right] }{  E^3_{\Delta^{(0)}, \boldsymbol{k}}(t_0) } 
           \right\}   ~  ,
           \label{Equation Delta1} 
\end{align}
\end{widetext}
where
\begin{align}
X_{ \Delta^{(0)}}(t)   \equiv     \frac{U}{2}   \sum_{\boldsymbol{k}}   \frac{ w_{\boldsymbol{k}} }{ E_{\Delta^{(0)}, \boldsymbol{k}}(t)    }  
       \left[ 1 - \frac{    \left| \Delta^{(0)}(t)  \right|^2 e^{i \phi^{(0)}(t)} }{  E^2_{\Delta^{(0)}, \boldsymbol{k}}(t)      }  \right]
\label{X_Delta}
\end{align}
and
\begin{align}
 \theta_{\Delta, \boldsymbol{k}}(t)   \equiv   \int_{t_0}^{t} \mathrm{d} t'  \left[
          \frac{       \epsilon_{\boldsymbol{k}} + f(t')     }{  E_{\Delta, \boldsymbol{k}}(t')   }      \, \partial_{t'} \phi(t')  - 2    E_{\Delta, \boldsymbol{k}}(t') \right]~.
          \label{alpha theta}
\end{align}

It should be noted that Eq.~\eqref{Equation Delta1} exhibits a typical feature of APT in that the corrections of higher order can be calculated from the knowledge of terms of lower orders only. So, once $\Delta^{(0)}(t)$ is known, the calculation of $\Delta^{(1)}(t)$ is numerically trivial. Concerning the determination of $\Delta^{(0)}(t)$, we see that the complexity of Eq.~\eqref{Equation Delta0} is comparable to that of the equilibrium BCS gap equation, except that the calculation should be done at each instant of time (on a grid). This is a minimal increase of computational complexity, which was expected in going from an equilibrium problem to the corresponding nonequilibrium one. A simplification of the equations is obtained by noticing that Eq.~\eqref{Equation Delta0} only determines the modulus of $\Delta^{(0)}(t)$. This quantity can be therefore chosen to be real. This sets the second line of Eq.~\eqref{Equation Delta1} to zero. However, $\Delta^{(1)}(t)$ develops an imaginary part.

Because $\theta_{\Delta^{(0)}, \boldsymbol{k}}(t_0) = 0$, one can verify that $\Delta^{(1)}(t_0) = 0$. Therefore, $\Delta^{(0)}(t_0)$ coincides with the total superconducting gap of the system at equilibrium, $\Delta(t_0) \equiv \Delta_0$.  

Finally, we also notice that $\Delta^{(0)}(t) \equiv 0  \,\, \forall t$ (which implies $\Delta^{(1)}(t) \equiv 0$ as well), corresponding to the normal state, is a possible solution.

\subsection{Constraints on the instantaneous range of variation of the zero-order APT gap  }
\label{subsec: rough estimate}

Discarding the normal-state solution, Eq.~\eqref{Equation Delta0} can be written as
\begin{align}
-      \frac{2}{U}   =    \sum_{\boldsymbol{k}}   \frac{ w_{\boldsymbol{k}} }{ E_{\Delta^{(0)}, \boldsymbol{k}}(t)   }  ~  .
\label{Equation Delta0 supercond}
\end{align}
This condition must be satisfied at all times $t$. Let us take Eq.~\eqref{Equation Delta0 supercond} and subtract the same Equation taken at $t = t_0$, using the fact that $f(t_0) = 0$. We obtain
\begin{align}
& 0  =    \sum_{\boldsymbol{k}} w_{\boldsymbol{k}}   \frac{ E_{\Delta_0, \boldsymbol{k}}(t_0) - E_{\Delta^{(0)}, \boldsymbol{k}}(t) }{ E_{\Delta^{(0)}, \boldsymbol{k}}(t) \, E_{\Delta_0 , \boldsymbol{k}}(t_0)   }   \nonumber \\
& =  \sum_{\boldsymbol{k}} w_{\boldsymbol{k}}   \frac{ E^2_{\Delta_0, \boldsymbol{k}}(t_0) - E^2_{\Delta^{(0)}, \boldsymbol{k}}(t) }{ E_{\Delta^{(0)}, \boldsymbol{k}}(t) \, E_{\Delta_0, \boldsymbol{k}}(t_0)  \left[ E_{\Delta_0, \boldsymbol{k}}(t_0) + E_{\Delta^{(0)}, \boldsymbol{k}}(t) \right]  }   \nonumber \\
& = c(t)   \left[ \left| \Delta_0 \right|^2   - f^2(t) - \left| \Delta^{(0)}(t) \right|^2 \right]  - 2 f(t) \, e(t) ~ ,
\label{estimate}
\end{align}
where we have defined
\begin{align}
& c(t) \equiv  \sum_{\boldsymbol{k}}    \frac{ w_{\boldsymbol{k}}  }{ E_{\Delta^{(0)}, \boldsymbol{k}}(t) \, E_{\Delta_0, \boldsymbol{k}}(t_0)  \left[ E_{\Delta_0, \boldsymbol{k}}(t_0) + E_{\Delta^{(0)}, \boldsymbol{k}}(t) \right]  }  ~, \nonumber \\
& e(t) \equiv  \sum_{\boldsymbol{k}}    \frac{ w_{\boldsymbol{k}} \epsilon_{\boldsymbol{k}} }{ E_{\Delta^{(0)}, \boldsymbol{k}}(t) \, E_{\Delta_0, \boldsymbol{k}}(t_0)  \left[ E_{\Delta_0 , \boldsymbol{k}}(t_0) + E_{\Delta^{(0)}, \boldsymbol{k}}(t) \right]}~.
\end{align}
Eq.~\eqref{estimate} would look like a very simple relation between the quantities $f(t)$, $\Delta^{(0)}(t)$ and $\Delta_0$, if it were not for the fact that $c(t)$ and $e(t)$ depend on those quantities as well. However, their ratio is a weighted sum of the quantities $\epsilon_{\boldsymbol{k}}$, which satisfies
\begin{align}
E_{1} \leq \frac{e(t)}{c(t)} \leq E_{2} \,\, \forall t~,
\end{align}
where $E_{1}$ and $E_{2}$ are the endpoints of the energy range introduced above in Sect.~\ref{subsec: external} in the Hubbard-BCS approximation. Setting $E_{2} \equiv E_{\rm D} - C$ and $E_{1} \equiv - E_{\rm D} - C$,  it follows that, for a given field $f(t)$, the following chain of inequalities must be satisfied:
\begin{align}
 & \left| \Delta_0 \right|^2   +  \left| f(t) \right| \! \Big[ 2   C \mathrm{sign} \left[f(t) \right]  - \left| f(t) \right|  -  2 E_{\rm D}   \Big] \leq \left| \Delta^{(0)}(t) \right|^2  \nonumber \\
 & \leq \left| \Delta_0 \right|^2   +   \left| f(t) \right| \Big[ 2   C \mathrm{sign} \left[f(t) \right]  - \left| f(t) \right|  + 2 E_{\rm D} \Big]~.
\label{range Delta}
\end{align}
This gives an {\it exact} (albeit not tight) constraint on how much $\Delta^{(0)}(t)$ can vary with respect to the equilibrium value $\Delta_0$ at each time $t$. 

	For example, Eq.~\eqref{range Delta} puts a restriction on the possibility to turn a normal material ($\Delta_0 = 0$) into a superconductor. In fact, for $\Delta_0 = 0$, Eq.~\eqref{range Delta} reduces to
\begin{align}
 &   \left| f(t) \right| \! \Big[ 2   C \mathrm{sign} \left[f(t) \right]  - \left| f(t) \right|  -  2 E_{\rm D}   \Big] \leq \left| \Delta^{(0)}(t) \right|^2  \nonumber \\
 & \leq   \left| f(t) \right| \Big[ 2   C \mathrm{sign} \left[f(t) \right]  - \left| f(t) \right|  + 2 E_{\rm D} \Big]~.
\label{normal - superc}
\end{align} 
Now, if $2   C \mathrm{sign} \left[f(t) \right]  - \left| f(t) \right|  + 2 E_{\rm D}  \leq 0$, Eq.~\eqref{normal - superc} admits no solutions or, in the case in which the equality applies, the solution is $\Delta^{(0)}(t) = 0$, implying that the system remains in the normal state, i.e.~the trivial solution of Eq.~\eqref{Equation Delta0} that was discarded in writing Eq.~\eqref{Equation Delta0 supercond}. This scenario cannot be altered by considering the additional term $\Delta^{(1)}(t)$ contributing to the dynamical gap for, as discussed earlier, $\Delta^{(0)}(t) = 0$ implies $\Delta^{(1)}(t) = 0$. 

So, in order to turn a normal material into a superconducting one, it is {\it necessary} (although not sufficient) that $2   C \mathrm{sign} \left[f(t) \right]  - \left| f(t) \right|  + 2 E_{\rm D} > 0$. 

The quantities
\begin{align}
E_{\rm D} \equiv (E_2 - E_1 ) / 2 > 0 ~, \quad C \equiv - (E_2 + E_1) / 2~,
\label{C and E_D}
\end{align}
which we have just introduced, depend on the specific system under consideration.

\subsection{The particular cases of initial thermal equilibrium and connection with the equilibrium case}
\label{subsec: recover equilibrium}

The present formulation, based on the KP contour, allows for a great flexibility in the choice of initial conditions~\cite{SecchiPolini1}. 

In this Section we check that the gap equation, in the case of an initial thermal superposition, reduces to the usual BCS gap equation at equilibrium. An initial thermal superposition corresponds to
\begin{align}
&  W_n = Z^{-1} \mathrm{e}^{- \beta \mathcal{E}_{n}(t_0)} = Z^{-1} \prod_{\boldsymbol{k}} \mathrm{e}^{- \beta  \left( n_{\boldsymbol{k}, +}  -  n_{\boldsymbol{k}, -} \right) E_{\boldsymbol{k}}(t_0)} ~, \nonumber \\
& Z = \sum_n W_n = \prod_{\boldsymbol{k}} \left[ 2 + \mathrm{e}^{- \beta E_{\boldsymbol{k}}(t_0)} + \mathrm{e}^{  \beta E_{\boldsymbol{k}}(t_0)} \right]~,
\end{align}
so that
\begin{align}
w_{\boldsymbol{k}} = \sum_n   W_n    \left( n_{\boldsymbol{k}, -} - n_{\boldsymbol{k}, +} \right)   = \tanh\left[ \beta E_{\boldsymbol{k}}(t_0) / 2 \right]~,
\end{align}
which should then be used to compute Eqs.~\eqref{Equation Delta0} and~\eqref{Equation Delta1}. For an initial thermal state and excluding the normal-state solution, Eq.~\eqref{Equation Delta0} reduces to
\begin{align}
 1 =    - \frac{U}{2}    \sum_{\boldsymbol{k}} \frac{  \tanh\left[ \beta \sqrt{\epsilon_{\boldsymbol{k}}^2 + \left| \Delta_0 \right|^2 }  / 2 \right]}{ \sqrt{\left[ \epsilon_{\boldsymbol{k}} + f(t) \right]^2 + \left| \Delta^{(0)}(t) \right|^2 } }~.
\label{equilibrium recovered}
\end{align}
In deriving the previous equation we assumed that $f(t_0) = 0$. If $f(t) = 0$, then $\Delta^{(0)}(t) = \Delta_0$ and $\Delta^{(1)}(t) = 0$, and Eq.~\eqref{equilibrium recovered} reduces to the standard BCS equation for the superconducting gap at equilibrium. In particular, it admits solutions only for $U < 0$ (attractive Hubbard model).

\section{Necessity and validity of Adiabatic Perturbation Theory}
\label{sec: necessity and validity}

\subsection{Inadequacy of the adiabatic theorem}
\label{subsec: time der}

We now discuss the problem that was anticipated in Sect.~\ref{sec: spatially uniform}, namely the inconsistency that emerges when using the adiabatic theorem of quantum mechanics, rather than the APT approach that we have pursued here.

Consider, in all generality, the case of a spatially non-uniform modulation of the gap parameter. This changes in time if $\dot{\Delta}_{\boldsymbol{q}}(t) \neq 0$. From Eq.~\eqref{mean-field equation} and using the Schr\"odinger equation satisfied by the states $\left| n(t) \right>$, we obtain
\begin{align}
 \dot{\Delta}_{\boldsymbol{q} }(t) &  = - i U \sum_n W_n    \left< n(t) \right| \sum_{\boldsymbol{k} } \left[ \hat{\mathcal{H}}(t) ,   \hat{d}^{\dagger}_{  \boldsymbol{k} + \boldsymbol{q}  , \downarrow  }  \hat{d}_{\boldsymbol{k},    \uparrow  } \right] \left| n(t) \right>  \nonumber \\
 &  =   i U  \sum_n W_n    \left\langle n(t) \right| \sum_{\boldsymbol{k} ,  \boldsymbol{k}'  }     \Big\{   \Delta_{\boldsymbol{k}' - \boldsymbol{k}}(t) \sum_{\sigma} \sigma \hat{d}^{\dagger}_{  \boldsymbol{k} + \boldsymbol{q}  , \sigma  }  \hat{d}_{\boldsymbol{k}',    \sigma  }    \nonumber \\
& \quad    + \! \left[  2 f_{\boldsymbol{k} - \boldsymbol{k}'}(t)  
 + \delta_{\boldsymbol{k},\boldsymbol{k}'} \! \left( \epsilon_{\boldsymbol{k}, 
 \uparrow} + \epsilon_{- \boldsymbol{k} - \boldsymbol{q}, \downarrow} \right) 
 \right] \hat{d}^{\dagger}_{  \boldsymbol{k} + \boldsymbol{q}  , \downarrow  }  
 \hat{d}_{\boldsymbol{k}',    \uparrow  } \! \Big\} \nonumber \\
 & \quad \times \left| n(t) \right\rangle~.
\end{align}
Let us restrict the analysis to the uniform case discussed above, with $f_{\boldsymbol{q}}(t)  = \delta_{\boldsymbol{q}, \boldsymbol{0}} f(t)$ and $\epsilon_{\sigma \boldsymbol{k}, \sigma} = \epsilon_{\boldsymbol{k}}$, and assume that $\Delta_{\boldsymbol{q}}(t) = \delta_{\boldsymbol{q}, \boldsymbol{0}} \Delta(t)$. We first need to check whether these two assumptions are consistent. If we assume so, the gap equation becomes
\begin{align}
 \dot{\Delta}(t) 
&  =   i U   \sum_n W_n \left< n(t) \right| \sum_{\boldsymbol{k}   } \Big\{    \Delta(t) \sum_{\sigma} \sigma \hat{d}^{\dagger}_{  \boldsymbol{k} , \sigma  }  \hat{d}_{\boldsymbol{k},    \sigma  } \nonumber \\
& \quad  +   2 \left[    f(t)  +    \epsilon_{\boldsymbol{k} }       \right] \hat{d}^{\dagger}_{  \boldsymbol{k}    , \downarrow  }  \hat{d}_{\boldsymbol{k},    \uparrow  } \Big\}   \left| n(t) \right>  ~, 
\label{derivative gap}
\end{align}
while the condition
\begin{align}
 0 &  =     \sum_n W_n \left< n(t) \right|     \sum_{\boldsymbol{k}    } \Big\{   \Delta(t) \sum_{\sigma} \sigma \hat{d}^{\dagger}_{  \boldsymbol{k} + \boldsymbol{q}  , \sigma  }  \hat{d}_{\boldsymbol{k},    \sigma  } \nonumber \\
& \quad +   \left[  2 f(t)  +    \epsilon_{\boldsymbol{k} } + \epsilon_{  \boldsymbol{k} + \boldsymbol{q} }   \right] \hat{d}^{\dagger}_{  \boldsymbol{k} + \boldsymbol{q}  , \downarrow  }  \hat{d}_{\boldsymbol{k},    \uparrow  } \Big\}  \left| n(t) \right>  ~, \quad \forall \boldsymbol{q} \neq \boldsymbol{0} 
 \label{consistency condition}
\end{align}
must be satisfied if the hypothesis that $\Delta_{\boldsymbol{q}}(t) = \delta_{\boldsymbol{q}, \boldsymbol{0}} \Delta(t)$ is valid. Using Eqs.~\eqref{inverse D d} we write these relations in terms of the $\hat{D}$ operators. Eq.~\eqref{consistency condition} becomes
\begin{align}
 0  & =         \sum_{\boldsymbol{k}' \neq \boldsymbol{k}   }  \sum_{\alpha, \alpha'} \Big\{   \Delta(t) \left[ a^*_{\boldsymbol{k} ,  \alpha}(t)  a_{\boldsymbol{k}', \alpha'}(t)  - b^*_{\boldsymbol{k} ,  \alpha}(t) b_{\boldsymbol{k}', \alpha'}(t)  \right]  \nonumber \\
& \quad +   \left[  2 f(t)  +    \epsilon_{\boldsymbol{k} } + \epsilon_{ \boldsymbol{k}'  }   \right] b^*_{\boldsymbol{k},  \alpha}(t)    a_{\boldsymbol{k}', \alpha'}(t)   \Big\} \nonumber \\
 &   \quad \times
 \sum_n W_n \left< n(t) \right|  \hat{D}^{\dagger}_{\boldsymbol{k} , \alpha}(t)  \hat{D}_{\boldsymbol{k}' , \alpha'}(t) \left| n(t) \right>   ~.
\end{align}
Under our assumptions, the total crystal momentum is a good quantum number, so
\begin{align}
\left< n(t) \right|  \hat{D}^{\dagger}_{\boldsymbol{k} , \alpha}(t)  \hat{D}_{\boldsymbol{k}' , \alpha'}(t) \left| n(t) \right>  = 0 ~, \quad \mathrm{if} \,\, \boldsymbol{k}' \neq \boldsymbol{k} ~,
\end{align}
and therefore Eq.~\eqref{consistency condition} is satisfied. Therefore, the assumptions that $\Delta_{\boldsymbol{q}}(t) = \delta_{\boldsymbol{q}, \boldsymbol{0}} \Delta(t)$ and $f_{\boldsymbol{q}}(t) = \delta_{\boldsymbol{q}, \boldsymbol{0}} f(t)$ are consistent. The same treatment applied to Eq.~\eqref{derivative gap} yields
\begin{align}
  \dot{\Delta}(t) 
 & = 
   i U  \mathrm{e}^{i \phi(t) } \sum_{\boldsymbol{k}   } \sum_{\alpha } \left\{          
     \alpha  \left[  \epsilon_{\boldsymbol{k} } +  f(t)   \right]        -     E_{\boldsymbol{k}}(t)     \right\}  \nonumber \\
     & \quad \times \sum_n W_n \left\langle n(t) \right| \hat{D}^{\dagger}_{\boldsymbol{k}, \alpha}(t) \hat{D}_{\boldsymbol{k}, - \alpha }(t) \left| n(t) \right\rangle~.
     \label{derivative Delta}
\end{align}
In the case of the adiabatic theorem, the states $\left| n(t) \right>$ appearing in Eq.~\eqref{derivative Delta} would be approximated, apart from phase factors, with the instantaneous eigenstates of $\hat{\mathcal{H}}(t)$, see Eq.~\eqref{instantaneous eigenstates}. Since these states have well-defined occupation numbers in the representation of the IQP fields, the bra-kets in the second line of Eq.~\eqref{derivative Delta} would vanish, yielding $\dot{\Delta}(t) = 0$ $\forall t$. This is in contradiction with the time-dependent solution of the dynamical gap equation coming from the adiabatic theorem. This shows why we could not have applied the adiabatic theorem for the treatment of our nonequilibrium problem. First-order APT gives a more accurate approximation of $\left| n(t) \right>$, including terms that do not conserve the IQP occupation numbers. This yields $\dot{\Delta}(t) \neq 0$, thereby removing the inconsistency.

\subsection{Applicability of first-order APT}
\label{subsec: applicability}

We now give a criterion to evaluate the accuracy of first-order APT, which we have used in this work. As mentioned in Sect.~\ref{subsec: APT}, in general one should require $\left| M_{n, m}(s) \right| \ll 1$. However, the most accurate condition obviously depends on the order of truncation of the APT expressions. In our case, we have approximated the time-dependent state as
\begin{align}
| \Psi(s) \rangle \approx | \Psi^{(0)}(s) \rangle + T^{-1} | \Psi^{(1)}(s) \rangle~.  
\end{align}
Let us compute the norm of this state using Eqs.~\eqref{Psi0 final} and~\eqref{Psi1 final}. We observe that
\begin{align}
\langle\Psi^{(0)}(s)| \Psi^{(0)}(s) \rangle = 1~,
\end{align}
\begin{align}
& \langle\Psi^{(0)}(s)| \Psi^{(1)}(s) \rangle =      i \sum_{  n \neq \Psi_0 }     J_{n , \Psi_0}(s)  \nonumber \\  
& = - \langle \Psi^{(1)}(s)| \Psi^{(0)}(s) \rangle~,
\end{align}
so that
\begin{align}
\langle\Psi(s)| \Psi(s) \rangle \approx 1 + T^{-2} \langle \Psi^{(1)}(s)| \Psi^{(1)}(s) \rangle~.
\end{align}
We then see that the condition
\begin{align}\label{APT-1 condition}
T^{-2} \langle \Psi^{(1)}(s)| \Psi^{(1)}(s) \rangle \ll 1
\end{align}
is a good test of the validity of first-order APT. In fact, this ensures an (approximate) instantaneous normalization of the time-dependent state, as well as it states that quantities which are formally of order $T^{-2}$ must be negligible. Some algebra is required in order to write Eq.~\eqref{APT-1 condition} explicitly. Specifically, one needs to use Eq.~\eqref{Psi1 final} and the explicit formulas for all the quantities appearing therein, which are given in Appendix~\ref{app: APT details}. It is convenient to introduce the set $\mathcal{S}_{\Psi_0} \equiv \left\{ (\boldsymbol{k}, \alpha ): n_{\boldsymbol{k}, \alpha} = 1 , n_{\boldsymbol{k}, - \alpha} = 0  \right\}$, where the quantities $ n_{\boldsymbol{k}, \alpha}$ are the quasi-particle occupation numbers characterizing the initial state $| \Psi_0 \rangle$ (see Section~\ref{subsect: Instantaneous}). The result for the left-hand side of Eq.~\eqref{APT-1 condition} then reads as
\begin{align}\label{condition APT-1 computational}
& T^{-2}\langle \Psi^{(1)}(s)| \Psi^{(1)}(s) \rangle \nonumber \\
&  =       \frac{1}{4 T^2}    \sum_{( \boldsymbol{k}, \alpha ) \in \mathcal{S}_{\Psi_0} }           \left|  \frac{   A_{\boldsymbol{k}}(s)_{ - \alpha ,  \alpha} \mathrm{e}^{- i  \alpha \theta_{\boldsymbol{k}}(s) }   }{   E_{\boldsymbol{k}}(s)    }           
            - \frac{   A_{\boldsymbol{k}}(s_0)_{ - \alpha ,  \alpha}   }{  E_{\boldsymbol{k}}(s_0)    }            \right|^2 \nonumber \\
        &   \quad + \frac{1}{4 T^2}  \left(  \sum_{( \boldsymbol{k}, \alpha ) \in \mathcal{S}_{\Psi_0} }   \int^s_{s_0} \mathrm{d} s'  \frac{ \left|  A_{\boldsymbol{k}}(s')_{ - \alpha  ,  \alpha}  \right|^2 }{ E_{\boldsymbol{k}(s')  } }\right)^2~,
\end{align}
where we have also used Eq.~\eqref{alpha theta}, and the quantity
\begin{align}
 A_{\boldsymbol{k}}(s)_{-\alpha, \alpha} =   \left[\partial_s {a}^*_{\boldsymbol{k}, -\alpha}(s) \right] a_{\boldsymbol{k}, \alpha}(s) + \left[\partial_s b^*_{\boldsymbol{k}, -\alpha}(s) \right] b_{\boldsymbol{k}, \alpha}(s) 
 \label{A with partial_s}
\end{align}
(see Appendix \ref{app: TD quasiparticle} for more details). It is intended that Eq.~\eqref{condition APT-1 computational} should be evaluated with $\Delta(t) \to \Delta^{(0)}(t)$. Importantly, since $\partial_s = T \partial_t$, one can easily see, after plugging Eq.~\eqref{A with partial_s} into Eq.~\eqref{condition APT-1 computational}, that the quantity \eqref{condition APT-1 computational} is independent of $T$.  

The check of whether the right-hand side of Eq.~\eqref{condition APT-1 computational} is $\ll 1$ should be carried out numerically, case by case. However, some simplifications occur in some relevant cases. First, if the initial state $\left| \Psi_0 \right>$ is the ground state, from Eq.~\eqref{instantaneous spectrum} we see that it must have $n_{\bm{k}, -1} = 1$ and $n_{\bm{k}, +1} = 0$, $\forall \boldsymbol{k}$. Therefore, in this case $\mathcal{S}_{\Psi_0} = \left\{   (\boldsymbol{k}, - )   \right\}$, and in Eq.~\eqref{condition APT-1 computational}, $\sum_{( \boldsymbol{k}, \alpha ) \in \mathcal{S}_{\Psi_0} }        \rightarrow \sum_{ \boldsymbol{k}  }    \sum_{\alpha} \delta_{\alpha, -}  $~.  

Then, under the assumptions that $\Delta^{(0)}(s)$ is real and $\neq 0$, one can obtain a relatively simple expression for Eq.~\eqref{A with partial_s} evaluated at $\Delta \rightarrow \Delta^{(0)}$ (see Appendix \ref{app: simplifications validity}): 
\begin{align}
 A_{\Delta^{(0)}, \boldsymbol{k}}(s)_{ - \alpha ,  \alpha} & =  \frac{\alpha \, \partial_s f(s)}{2  \left| \Delta^{(0)}(s)  \right| E^2_{\Delta^{(0)}, \boldsymbol{k}}(s) } \Bigg\{ \left| \Delta^{(0)}(s)  \right|^2  \nonumber \\
& \quad + \left[ J_{\Delta^{(0)}}(s) + f(s) \right] \left[ \epsilon_{\boldsymbol{k}} + f(s) \right]   \Bigg\} ~, 
\label{after algebra}
\end{align}
where
\begin{align}\label{J}
J_{\Delta^{(0)}}(s) \equiv \left( \sum_{\boldsymbol{k}}    \frac{ w_{\boldsymbol{k}}   \epsilon_{\boldsymbol{k}}   }{ E^3_{\Delta^{(0)}, \boldsymbol{k}}(s)   } \right) \Bigg/ \left( \sum_{\boldsymbol{k}}    \frac{ w_{\boldsymbol{k}}   }{ E^3_{\Delta^{(0)}, \boldsymbol{k}}(s)   } \right)~.
\end{align}
Eq.~\eqref{after algebra} can then be inserted into Eq.~\eqref{condition APT-1 computational} before numerical evaluation. Since this is system-dependent, such numerical analysis is well beyond the scope of the general theoretical framework that we are formulating here.

\section{Analytical results at zero temperature}
\label{sec: example}

In this Section, we obtain {\it the} analytical expressions for solutions of the problem at hand, i.e.~for the zero- and first-order APT components of the nonequilibrium gap, in the case of initial equilibrium and zero temperature ($\beta \rightarrow \infty$).

As mentioned in Sect.~\ref{sec: Hamiltonian}, we suppose that the superconducting mechanism is due to the interaction between electrons and one branch of RA phonons. The applied electromagnetic field activates a $\boldsymbol{q} = \boldsymbol{0}$ IRA phonon, which is coupled to the $\boldsymbol{q} = \boldsymbol{0}$ RA phonon through a type-I phonon nonlinearity. So, the phonons involved in our considerations are optical. We will not specify a particular material, but rather consider general trends at zero temperature, leaving the study of the case of finite temperature, as well as the application to specific systems, to future works.

\subsection{Adiabatic term of the superconducting gap}

The adiabatic term $\Delta^{(0)}(t)$ is obtained from Eq.~\eqref{equilibrium recovered}, which we write in terms of an integral on the electronic effective energies,
\begin{align}
- \frac{2}{U}   =   \int_{E_1}^{E_2} d \epsilon  \sigma(\epsilon) \frac{\tanh\left[ \frac{\beta}{2} \sqrt{    \epsilon^2 + \left| \Delta_0 \right|^2 } \right] }{\sqrt{  \left[ \epsilon  + f(t) \right]^2 + \left| \Delta^{(0)}(t) \right|^2 }} ~,
\label{what we want}
\end{align}
where $\sigma(\epsilon)$ is the density of states, and the integration limits $E_1$ and $E_2$ have been introduced in Sect.~\ref{subsec: HS}. It is customary in BCS theory to assume that $\sigma(\epsilon) \approx {\rm const.} \equiv \sigma_0$ in the range of integration. Adopting the same approximation, we can put
\begin{align}
\sum_{\bm{k}} F(\epsilon_{\bm{k}})   \approx \sigma_0 \int_{E_1}^{E_2}   d\epsilon   F(\epsilon )~,
\label{conversion k energy}
\end{align}
for all functions $F(\epsilon_{\boldsymbol{k}})$ that depend on $\boldsymbol{k}$ only through $\epsilon_{\boldsymbol{k}}$.  
 
If we replace $\sigma(\epsilon) \approx \sigma_0$ and let $\beta \rightarrow \infty$, the remaining integral in Eq.~\eqref{what we want} can be performed analytically:
\begin{align}
& \int_{E_1}^{E_2} d\epsilon \frac{1 }{\sqrt{  \left[ \epsilon + f(t) \right]^2 + \left| \Delta^{(0)}(t) \right|^2 }} \nonumber \\
& = \ln \left(\frac{E_2  + f(t) + \sqrt{    \left[ E_2 + f(t) \right]^2 + \left| \Delta^{(0)}(t) \right|^2 }   }{E_1  + f(t) + \sqrt{    \left[ E_1 + f(t) \right]^2 + \left| \Delta^{(0)}(t) \right|^2 }} \right) ~,
\end{align}
and Eq.~\eqref{what we want} becomes
\begin{align}
\frac{E_2  + f(t) + \sqrt{    \left[ E_2 + f(t) \right]^2 + \left| \Delta^{(0)}(t) \right|^2 }   }{E_1  + f(t) + \sqrt{    \left[ E_1 + f(t) \right]^2 + \left| \Delta^{(0)}(t) \right|^2 }} = x ~,
\label{equation to be solved}
\end{align} 
where
\begin{align}
x \equiv  \exp\left( \frac{2  }{- U \sigma_0} \right) ~.
\label{parameter x}
\end{align}
Since $U < 0$, we have $x > 1$.

\subsection{Gap as a function of the applied field}

We  now solve Eq.~\eqref{equation to be solved} for $\left| \Delta^{(0)}(t) \right|$. It is convenient to express the result in terms of the quantities $C$ and $E_{\rm D}$ introduced in Eq.~\eqref{C and E_D}. We find
\begin{align}\label{nonequilibrium gap ED C}
\left| \Delta^{(0)}(t) \right|   =     \sqrt{ \left| \Delta_0 \right|^2  + \frac{4x}{(x+1)^2} \left[  2 C f(t) - f^2(t)   \right]}~,
\end{align}
where the equilibrium gap $\left| \Delta_0 \right|$ is
\begin{align}\label{equilibrium gap ED C}
\left| \Delta_0 \right|   =   2 \sqrt{x} \sqrt{ \frac{    E_{\rm D}^2}{(x-1)^2} - \frac{  C^2}{(x+1)^2}}~.
\end{align}
It should be noted that $E_{\rm D}$ is not the Debye energy, but rather the band width of the branch of RA phonons in our model, and it is equal to half the width of the integration range in Eq.~\eqref{what we want} (independently of $C$). Moreover, we note that conventional (equilibrium) BCS theory postulates $C = 0$. In this case, the range of integration in Eq.~\eqref{what we want} is centered on $\epsilon = 0$, where $\epsilon \rightarrow \epsilon_{\boldsymbol{k}} = \epsilon^{(0)}_{\boldsymbol{k}} - \mu$. However, this assumption has been criticized~\cite{Anghel16}, because the electron-phonon interaction is a microscopic feature of the system and, as such, it should not be so directly tied to the value of the electronic chemical potential $\mu$, which can be changed by applied pressure or doping. Therefore, the authors of Ref.~\onlinecite{Anghel16} considered the possibility of centering the integration range on a different effective chemical potential than the electronic one, an approach that is equivalent to taking $C \neq 0$ in our formalism. It was shown there that, at equilibrium, $C \neq 0$ gives remarkable differences with respect to the $C = 0$ case postulated in standard BCS, including the fact that the phase transition becomes of the first order (while being of the second order only if $C = 0$).

In our case, we recall that, after the Nambu transformation, we had obtained (compare with Eq.~\eqref{effective Nambu band})
\begin{equation}
\epsilon_{  \boldsymbol{k} } \equiv \epsilon^{(0)}_{  \boldsymbol{k} }  - \mu   - 2 \mathcal{N}  \sum_{  \lambda}          (    M^2_{\boldsymbol{0}, \lambda} \, / \, \omega_{\boldsymbol{0}, \lambda}   ) ~. 
\end{equation} 
The last constant in the right-hand side of this expression (which might be $\neq 0$ for optical phonons) comes from merely algebraic steps after the Nambu substitution, so it should not alter the physical considerations on which BCS is based. If we accept the BCS assumption, the quantity $\epsilon_{\boldsymbol{k}}^{(0)} - \mu$ is restricted to 
\begin{align}
- E_{\rm D} < \epsilon_{\boldsymbol{k}}^{(0)} - \mu < E_{\rm D} ~.
\end{align} 
However, the integration variable in Eq.~\eqref{what we want} is not $\epsilon_{\boldsymbol{k}}^{(0)} - \mu$, but $\epsilon_{\boldsymbol{k}}$, which in our case lies instead in the interval
\begin{align}
- E_{\rm D} - 2 \mathcal{N} \sum_{\lambda} \frac{M^2_{\boldsymbol{0}, \lambda}}{\omega_{\boldsymbol{0}, \lambda}} < \epsilon_{\boldsymbol{k}}  < E_{\rm D} - 2 \mathcal{N} \sum_{\lambda} \frac{M^2_{\boldsymbol{0}, \lambda}}{\omega_{\boldsymbol{0}, \lambda}} ~.
\end{align} 
This suggests the identification
\begin{align}
C \equiv   2 \mathcal{N} \sum_{\lambda} \frac{M^2_{\boldsymbol{0}, \lambda}}{\omega_{\boldsymbol{0}, \lambda}}  > 0 
\end{align}
(the sum over $\lambda$ is actually restricted to the single RA branch under consideration).

So, in the following we take $C \geq 0$ as a material-specific parameter. We immediately see that $C \neq 0$ leads to a number of remarkably interesting features:
\begin{itemize}
\item[1)] The equilibrium gap at $\beta \rightarrow \infty$ (see Eq.~\eqref{equilibrium gap ED C}) exists only if 
\begin{align}
C < E_{\rm D} \frac{x+1}{x-1} \equiv C_{\rm max} 
\label{existence equilibrium gap}
\end{align}
(we recall that $E_{\rm D} > 0$, $C \geq 0$, $x > 1$), so it is not guaranteed that the superconducting phase exists at zero temperature (which would be the case for $C = 0$). Considering $\left| \Delta_0 \right| $ as a function of $C$, its maximum value is achieved at $C = 0$, in agreement with the result of Ref.~\onlinecite{Anghel16}.

\item[2)] The nonequilibrium gap (see Eq.~\eqref{nonequilibrium gap ED C}) exists only if
\begin{align}
C - \frac{  x+1 }{ x-1  }     E_{\rm D}
 < f(t) < 
C + \frac{  x+1 }{ x-1  }     E_{\rm D}~,
\label{constraint on f}
\end{align}
where we have used Eq.~\eqref{equilibrium gap ED C}. If $f(t)$ falls out of this range at a given $t$, the superconducting phase is destroyed.  

\item[3)] The nonequilibrium gap at time $t$ is {\it enhanced} with respect to the equilibrium gap if
\begin{align}
2 C f(t) - f^2(t) > 0 \Rightarrow 0 < f(t) < 2 C~,
\end{align}
which is valid for $C > 0$ (otherwise, if $C < 0$, the condition would be $2 C < f(t) < 0$). Note that, if $C = 0$, it is not possible to increase the superconducting gap with respect to the equilibrium value (at least, at zero temperature and with the proposed mechanism), although it is possible to modulate it in time. Therefore, $C$ plays a crucial role in our model.

\item[4)] Provided that $2 C f(t) - f^2(t) > 0$, it is possible to have $\left| \Delta_0 \right| = 0$ and $\left| \Delta^{(0)}(t) \right| > 0$, i.e., to trigger a transient superconducting state starting from a normal state at equilibrium.

\item[5)] After rewriting Eq.~\eqref{nonequilibrium gap ED C}, combined with \eqref{equilibrium gap ED C}, as
\begin{align}
\left| \Delta^{(0)}(t) \right|   =    2 \sqrt{x}  \sqrt{ \frac{    E_{\rm D}^2}{(x-1)^2} - \frac{ \left[ f(t)  -   C \right]^2}{(x+1)^2}}~,
\label{obvious maximization}
\end{align}  
we immediately see that $\left| \Delta^{(0)}(t) \right|$ takes it maximum possible value when $f(t) = C$. The maximum value is
\begin{align}
\left| \Delta^{(0)}_{\rm max} \right|        =        \frac{  2 \sqrt{x}   }{ x-1  } E_{\rm D}~.
\end{align}
Therefore the applied field can, at most, cancel the lowering effect of $C \neq 0$ on the equilibrium gap (see Eq.~\eqref{equilibrium gap ED C}). If $\left| \Delta_0 \right| \neq 0$ (therefore, $C < C_{\rm max}$), the maximum relative increase of the gap is
\begin{align}
\frac{ \left| \Delta_{\rm max}^{(0)}  \right| }{\left| \Delta_0 \right|} = \left(      1  - \frac{    C^2}{C_{\rm max}^2}         \right)^{-1/2}~.
\label{max Delta}
\end{align}

\end{itemize}

\subsection{Applied field needed to obtain the desired gap}

We now consider the inverse problem. Imagine one desires that $\left| \Delta^{(0)}(t) \right|$ is a specific function of time and we have to find the applied field $f(t)$ that generates it. To obtain the desired expression, we solve Eq.~\eqref{nonequilibrium gap ED C} for $f(t)$, obtaining the two solutions
\begin{align}
  f_{\pm}(t) = C \pm  (x+1) \sqrt{     \frac{E_{\rm D}^2 }{ (x-1)^2 }     - \frac{1}{4x} \left| \Delta^{(0)}(t) \right|^2}~.
\label{time-dependent applied field}
\end{align} 
Since $f(t) \in \mathbb{R}$, the condition for the solutions to exist is that $ \left| \Delta^{(0)}(t)  \right|  <  \left| \Delta_{\rm max}^{(0)}  \right| $, consistently with the discussion in the previous Section.

\subsection{The effective external field: explicit expression}
\label{subsec: ex explicit f}

Keeping only the $\boldsymbol{q} = \boldsymbol{0}$ term, we re-write the Hamiltonian for the external field, Eq.~\eqref{external field}, as
\begin{align}
\hat{\mathcal{H}}_{\mathrm{ext}}(t) =  F(t) \sqrt{2   \omega_{\boldsymbol{0}}^{ \mathrm{RA}}} \hat{Q}_{\boldsymbol{0}}^{\mathrm{RA}}~,
\label{external at q = 0}
\end{align}
where we have specified a RA phonon branch. For the sake of brevity, we write $\omega_{\boldsymbol{0}}^{ \mathrm{RA}} \equiv \omega$. Identifying Eq.~\eqref{external at q = 0} with the nonlinear coupling of type I between RA and IRA phonons (see Eq.~\eqref{I-nonlinearity term}), we set it equal to $\Lambda_{\rm I} \left[ Q_{\boldsymbol{0}}^{\mathrm{IRA}}(t)\right]^2 \hat{Q}_{\boldsymbol{0}}^{\mathrm{RA}}$, obtaining
\begin{align}
F(t) = \frac{\Lambda_{\rm I}}{\sqrt{2   \omega }} \left[ Q_{\boldsymbol{0}}^{\mathrm{IRA}}(t)\right]^2~.
\label{F_IRA}
\end{align}
Following Ref.~\onlinecite{Knap16} we now assume that $Q_{\boldsymbol{0}}^{\mathrm{IRA}}(t)$ is a trigonometric function of time and, in order to simulate the switch on of the field, we put
\begin{align}
Q_{\boldsymbol{0}}^{\mathrm{IRA}}(t) = \theta(t - t_0) \mathcal{Q} \sin\left[ \Omega  (t - t_0)\right]~,
\label{Q_IRA}
\end{align} 
where $\Omega$ is the pumping frequency of the applied field. We do not switch the field off, since we are interested in the transient dynamics and the leading APT term of the gap closely follows the time dependence of $f(t)$.

Inserting Eq.~\eqref{F_IRA} into Eq.~\eqref{f(t)}, we obtain 
\begin{align}
f(t)  
& =  -  \frac{2 M^{\mathrm{RA}}_{\mathbf{0}} \Lambda_{\rm I}}{\sqrt{2   \omega_{\mathbf{0}}^{ \mathrm{RA}}}}     \int_{- \infty}^t dt' \sin\! \left[ \omega_{\mathbf{0}}^{ \mathrm{RA}} (t - t') \right] \left[ Q_{\mathbf{0}}^{\mathrm{IRA}}(t)\right]^2~.
\label{f(t) as an integral}
\end{align}
Using Eq.~\eqref{Q_IRA}, the integral \eqref{f(t) as an integral} can be carried out analytically. Introducing the quantity 
\begin{align}
A \equiv -  \frac{  M^{\mathrm{RA}}_{\mathbf{0}} \Lambda_{\rm I} \mathcal{Q}^2 }{\omega \sqrt{2 M \omega   }}~,
\label{parameter A}
\end{align}
the result (for $t > t_0$) is 
\begin{align}
f(t) 
& =     A   \left\{ 1 - \frac{4 \Omega^2}{ 4 \Omega^2 - \omega^2 } \cos\! \left[ \omega  (t - t_0) \right] \right. \nonumber \\
& \quad \left. + \frac{  \omega^2}{ 4 \Omega^2 - \omega^2 } \cos\! \left[ 2 \Omega  (t - t_0) \right]      \right\}~.
\label{f(t) explicit}
\end{align}
The first time derivative is
\begin{align}
\dot{f}(t) 
& =     A  \frac{2 \Omega \omega}{  4 \Omega^2 - \omega^2  }  \Big\{  2 \Omega   \sin\! \left[ \omega  (t - t_0) \right]   \nonumber \\
& \quad   -      \omega   \sin\! \left[ 2 \Omega  (t - t_0) \right]    \Big\} ~.
\label{dot f(t) explicit} 
\end{align}
We have that $f(t_0) = 0$, as required in the derivation, and $\dot{f}(t_0) = 0$, which will lead to a significant simplification in the analytical expression for $\Delta^{(1)}(t)$---see Sect.~\ref{subsec: ex Delta1}.

\subsection{First-order APT correction to the gap}
\label{subsec: ex Delta1}

We now simplify Eq.~\eqref{Equation Delta1} for the case of zero temperature, i.e.~for $\beta \rightarrow \infty$. Assuming that $\Delta^{(0)}(t)$ is real and positive, we replace $\left| \Delta^{(0)}(t) \right|  \rightarrow \Delta^{(0)}(t) $ in Eq.~\eqref{nonequilibrium gap ED C} and notice that the whole second line of Eq.~\eqref{Equation Delta1} vanishes. Then, we notice that the whole third line of Eq.~\eqref{Equation Delta1} vanishes as well, because it is proportional to $\dot{f}(t_0)$---since $\dot{\Delta}^{(0)}(t_0) \propto \dot{f}(t_0)$---and we have seen in Section \ref{subsec: ex explicit f} that we can take $\dot{f}(t_0) = 0$. Then, we convert the summations over $\boldsymbol{k}$ into integrals over $d \epsilon$ according to Eq.~\eqref{conversion k energy} and, using the fact that $ w_{\boldsymbol{k}} =  \tanh\left[ \beta E_{\boldsymbol{k}}(t_0) / 2 \right] \rightarrow 1 $, we can carry out all the integrals analytically. We have laid down the most significant steps of the algebraic manipulations in Appendix~\ref{app: simplifications Delta_1}. We here state only the final result:
\begin{align}
  \Delta^{(1)}(t)  & =  -   i \, \frac{ \dot{f}(t) }{ 2     \Delta^{(0)}(t) } \left[ 1        +  \frac{16 x^2}{(x+1)^4}   \frac{ [ f(t) - C ]^2      }{       \left[ \Delta^{(0)}(t) \right]^2    }        \right]  ~.  
  \label{Equation Delta1 simplified} 
\end{align} 
Since $\Delta^{(1)}(t) $ is purely imaginary, while $\Delta^{(0)}(t) $ is real, the square modulus of the total gap
\begin{align}
| \Delta(t)|^2 \approx | \Delta^{(0)}(t)  + \Delta^{(1)}(t)|^2 = | \Delta^{(0)}(t)|^2 + |\Delta^{(1)}(t)|^2
\end{align}
is always larger or equal with respect to the adiabatic term alone (within first-order APT).

We notice that $\Delta^{(1)}(t)  \propto \dot{f}(t) $, so the first-order term of the gap vanishes at the stationary points of $\Delta^{(0)}(t)$, since $\dot{\Delta}^{(0)}(t) \propto \dot{f}(t)$. Therefore, in correspondence of those points (and in sufficiently small neighbourhoods enclosing them), the adiabatic term reliably accounts for the whole gap.

\subsection{Numerical examples}

We now report illustrative numerical results related to two relevant cases. In the following, we put $t_0 = 0$, and choose $\Delta_{0}$ as our unit of energy---see Eq.~(\ref{equilibrium gap ED C}). We plot the dimensionless quantities $f(t) / \Delta_{0}$, $\Delta^{(0)}(t) / \Delta_{0}$, and $|\Delta(t)| / \Delta_{0}$ as functions of the variable $\omega t / (2 \pi)$, for selected values of the input parameters $x$ and $C / E_{\rm D}$, and two different values of the pumping frequency $\Omega$.

\subsubsection{Intrinsic parameters of the superconducting system}

Having taken $\Delta_0$ as unit of energy, the only other parameters related to the superconducting system are $x$ and the ratio $C / E_{\rm D}$. The quantity $x$ is extremely sensitive to the model parameters $U$ and $\sigma_0$ (see Eq.~\eqref{parameter x}). According to Ref.~\onlinecite{Tinkham}, $- U \sigma_0 < 0.3$ for most classic superconductors. However, in our case, the superconducting behaviour is dictated by optical RA phonons, whose coupling strength with the electronic system ($-U$) is generally larger than the coupling between electrons and acoustic phonons. Note that a small change in $U$ has a huge repercussion on $x$. For example, in the weak-coupling regime ($- U \sigma_0 \ll 1$), taking e.g.~$- U \sigma_0 \approx 0.2$, we get
\begin{align}
x = e^{10} \approx 2.2 \times 10^4~,
\end{align} 
while in the intermediate regime, taking e.g.~$- U \sigma_0 \approx 0.96$, we get
\begin{align}
x = e^{2/0.96} \approx 8.0~.
\end{align}

We target the ideal situation in which  the nonequilibrium superconducting gap is significantly increased with respect to the equilibrium one and, at the same time, the first-order APT contribution to the gap is relatively small with respect to the adiabatic term. Within the parameter space where this occurs, we here focus for the sake of definiteness on the following specific choice: $x = 8$ and $C / E_{\rm D} = 0.9$~.

\subsubsection{Parameters of the external field}

With respect to the pumping frequency $\Omega$, we consider two relevant cases: $\Omega = 4 \omega$, and $\Omega = \omega / 4$. Choosing $\omega$ and $\Omega$ to be commensurate makes all the functions periodic, with period equal to $2 \pi / \min(\omega, \Omega)$, which simplifies the visualization. 

The motivation to study the case $\Omega = 4 \omega$ comes from the parameters of PMO, a paradigmatic material displaying type-I nonlinearity. In Ref.~\onlinecite{Subedi14}, it is found that $\omega^{\mathrm{RA}}_{\boldsymbol{0}} = 155$ cm$^{-1}$, while $\omega^{\mathrm{IRA}}_{\boldsymbol{0}} = 622$ cm$^{-1}$, so $\omega^{\mathrm{IRA}}_{\boldsymbol{0}} \approx 4 \omega^{\mathrm{RA}}_{\boldsymbol{0}}$ (we have retained their units). If the IRA phonon is pumped resonantly, i.e. $\Omega = \omega^{\mathrm{IRA}}_{\boldsymbol{0}}$, then we have almost exactly $\Omega \approx 4 \omega$. 

The opposite case $\Omega = \omega / 4$, representing a pumping well below resonance, turns out to be more suitable for an APT treatment and therefore, for the optimal control of the gap modulation. In fact, it is easy to see from Eq.~\eqref{dot f(t) explicit} that, when there is a large difference between $2 \Omega$ and $\omega$, one has 
\begin{align}
\max | \dot{f}(t)| \approx |A| \min(\omega, 2 \Omega) ~.
\end{align}
This value dictates the maximum absolute value of $\Delta^{(1)}(t)$, which should be small in view of APT. Therefore the second case, where $\min(\omega, 2 \Omega) = \omega / 2$, is better, from this point of view, than the first case, where $\min(\omega, 2 \Omega) = \omega$.

We then consider the amplitude $A$ in Eq.~\eqref{parameter A}. Since $f(t)$ must lie in the interval given by Eq.~\eqref{constraint on f}, otherwise the gap is destroyed, we fix $A$ in both cases to $0.25 A_{\rm max}$, where $A_{\rm max}$ is the critical value at which $f(t)$ periodically touches one of the boundaries of the interval given in Eq.~\eqref{constraint on f}. In this way, we are well below the critical amplitude, and the gap is well defined at all times.

\begin{figure}
 \includegraphics[width=\linewidth]{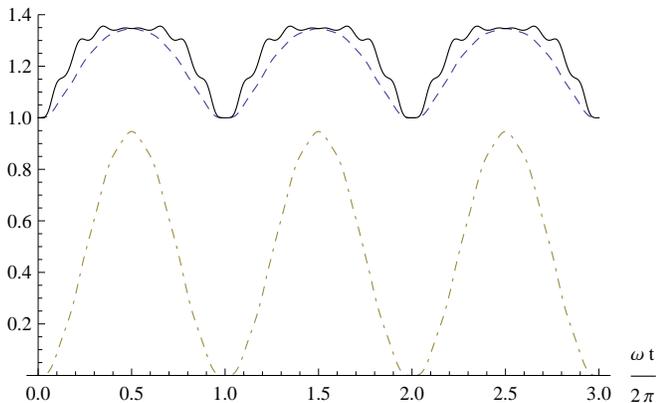}
  \caption{ (Color online) Plots of the dimensionless quantities $f(t) / \Delta_{0}$ [red (dash-dotted) curve], $\Delta^{(0)}(t) / \Delta_{0}$ [blue (dashed) curve], and $|\Delta(t)| / \Delta_{0}$ [black (solid) curve], as functions of $\omega t / (2 \pi)$. Numerical results in this plot have been obtained by setting $C = 0.9 E_{\rm D}$, $x = 8$, $\omega = 1.2 E_{\rm D}$, and $\Omega = 4 \omega$. The equilibrium gap $\Delta_{0}$ can be easily calculated from Eq.~(\ref{equilibrium gap ED C}). All the rescaled quantities displayed in this Figure do not depend separately on $C$ and $E_{\rm D}$ but only on the ratio $C/E_{\rm D}$. \label{fig: big Omega}}
\end{figure}
\begin{figure}
 \includegraphics[width=\linewidth]{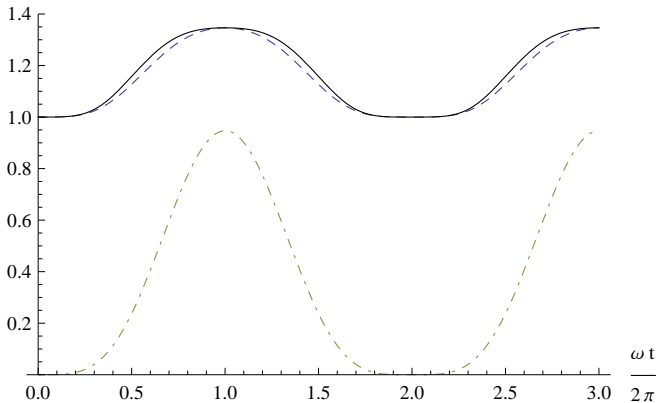}
  \caption{(Color online) Same as in Fig.~\ref{fig: big Omega}, but for $\Omega = \omega / 4$. \label{fig: small Omega}}
\end{figure}

\subsubsection{Numerical results}

Fig.~\ref{fig: big Omega} displays our main findings for a pumping frequency $\Omega = 4 \omega$. We see that the adiabatic component $\Delta^{(0)}(t)$ of the gap [blue (dashed) curve] closely follows the time dependence of $f(t)$ [red (dashed-dotted) curve]. The modulus of the total gap [black (solid) line], $|\Delta(t)|$, is slightly altered by the enhancing effect of the first-order APT term. Note that the value of $|\Delta(t)|$ at times $ t = (2 n + 1)\pi / \omega $ is $\approx 35\%$ larger than the equilibrium value $\Delta_{0}$, which is repeatedly reached periodically at times $ t =  2 n  \pi / \omega $.

Fig.~\ref{fig: small Omega} is related, instead, to the case $\Omega = \omega / 4$. We see that, qualitatively, the scenario is similar to that occurring in the high-pumping-frequency case. However, the correction to the total gap due to the first-order APT term is smaller, as well as much smoother, than in the high-pumping-frequency case. Of course, choosing a lower pumping frequency is much more in line with the idea underlying APT and, at the same time, it produces a gap enhancement that lasts for longer times (note that the time scale on the horizontal axes in Figs.~\ref{fig: big Omega} and \ref{fig: small Omega} is the same).

\section{Summary and conclusions}
\label{sec: conclusions}

In this Article we have laid down a theoretical framework to compute the nonequilibrium superconducting gap for a coupled electron-phonon system subject to an external time-dependent electromagnetic field acting on the phonon subsystem. Since our main objective was to transcend the limitations of Floquet theory and/or heavy numerical methods, we had to make an assumption of slow time dependence of the external field.  

As it happens with any approximate scheme, our formulation has both advantages and drawbacks with respect to previous works. From the point of view of the {\it model}, our approach has the advantage that it does not restrict the external field to be periodic (contrarily to Floquet theory), nor to have a small amplitude. The only restriction is that APT must be valid, which should be assessed case by case from the numerical evaluation of Eq.~\eqref{condition APT-1 computational}. One of the advantages of APT is that, in principle, further, higher-order corrections can be included, although the formulas get more involved. However, each additional correction can be computed from the knowledge of the lower-order terms contributing to $\Delta(t)$. Therefore, the most numerically-demanding task is the solution of Eq.~\eqref{Equation Delta0 supercond}.

From the {\it computational} point of view, our formulas exhibit the minimal increase of computational complexity that can be expected in going from an equilibrium to a nonequilibrium problem. The nonequilibrium problem is mapped onto a set of equilibrium-like problems to be solved at every instant of time on a grid. The required computation, being analogous to an equilibrium BCS one, does not exhibit numerical difficulties such as the determination of huge two-times Green's functions and self-energy matrices, which would require to adopt simple expressions for the time-dependent field. The drawback of our approach is that it is based on mean-field theory, which can be transcended if a full numerical Keldysh calculation is done as e.g. in Ref.~\onlinecite{Sentef16} (some effects that are not captured within mean-field theory are discussed there). It should be noted that the mean-field theory proposed in the Appendix of Ref.~\onlinecite{Sentef16} is still to be intended as a numerical approximation, i.e.~as a way to compare the full numerical calculation with a simpler one, but it is not equivalent to the semi-analytical formulas presented here [Eqs.~\eqref{Equation Delta0} and~\eqref{Equation Delta1}].

One of the earliest discussions on nonequilibrium superconductivity can be found in Ref.~\onlinecite{Abrahams66}. It treats the case of a time- and space-dependent gap parameter, and it derives differential equations for it under several conditions, by means of second-order Taylor expansions in space and time gradients. Such an approach presents several difficulties, which are thoroughly discussed in Ref.~\onlinecite{Abrahams66}. The resulting equations, depending on the various situations discussed, either assume smallness in the size variation of $\Delta$, or  are valid on short time intervals due to use of a Taylor expansion. Moreover, the results are differential equations whose numerical solution is demanding. The advantage of our APT-based approach is that it reduces the problem to the solution of {\it algebraic} equations, whose validity is not restricted to small variations of $\Delta$ nor to small time intervals, provided that the external field is slow.

Our analytical theory can be used to answer several intriguing questions. For example, one may consider the problem of whether a non-superconducting material, with $\Delta(t_0) = 0$, can be driven into a non-equilibrium superconducting state, with $\Delta(t) \neq 0$, by applying an external time-dependent electromagnetic field. 
From Eq.~\eqref{equilibrium recovered} one sees that these two requirements are compatible, and with Eq.~\eqref{normal - superc} we have provided a necessary (but not sufficient) condition for the transition to occur, which is aimed to guide computational studies. Since the normal state solution is always possible, it should be noted that the study of the normal-superconducting transition driven by the application of an external field requires a further stability analysis~\cite{Aronov73}.

Importantly, an explicit analytical solution of Eqs.~\eqref{Equation Delta0} and \eqref{Equation Delta1} at zero initial temperature is reported in Sect.~\ref{sec: example}, together with some illustrative numerical results---see Figs.~\ref{fig: big Omega}-\ref{fig: small Omega}.

\acknowledgements

This work was supported by the European Union's Horizon 2020 research and innovation programme under grant agreement No. $785219$ - ``GrapheneCore2''. We wish to thank Andrea Cavalleri and Andrea Tomadin for useful and inspiring discussions.

\appendix

\section{Bosonic Gaussian integration on the KP time contour}
\label{app: bosonic int}

We simplify Eq.~\eqref{Z KB} by carrying out the bosonic integral
\begin{align}\label{bosonic integral}
  \int \mathcal{D}(a^*, a) \mathrm{e}^{i  S_{\mathrm{ep}}}    = \prod_{\boldsymbol{q} , \lambda}   \int \mathcal{D}(a^*_{\boldsymbol{q} , \lambda}, a_{\boldsymbol{q} , \lambda})  \mathrm{e}^{ i S_{\mathrm{ep}; \boldsymbol{q} , \lambda}}~.
\end{align}
We re-write Eq.~\eqref{bosonic integral} as
\begin{align}
& \prod_{\boldsymbol{q}, \lambda}    \int \mathcal{D} \left( a^*_{\boldsymbol{q}, \lambda} , a_{\boldsymbol{q}, \lambda} \right)  \nonumber \\
& \times \exp\Bigg\{ i \iint_{\gamma} \mathrm{d}z \mathrm{d}z' \, a^*_{\boldsymbol{q}, \lambda}(z) \, \hat{G}^{\mathrm{fp} -1}_{ \boldsymbol{q}, \lambda}(z,z') \, a_{\boldsymbol{q}, \lambda}(z') \nonumber \\
&  - i \int_{\gamma} \mathrm{d}z \left[  \widetilde{J}_{\boldsymbol{q}, \lambda}(z) a_{\boldsymbol{q}, \lambda}(z) + a^*_{\boldsymbol{q}, \lambda}(z) J_{\boldsymbol{q}, \lambda}(z)  \right]    \Bigg\}  \nonumber \\
& = \prod_{\boldsymbol{q}, \lambda}  \frac{ \exp\left\{ - i \iint_{\gamma} \mathrm{d}z \mathrm{d}z'  \widetilde{J}_{\boldsymbol{q}, \lambda}(z) G^{\mathrm{fp}}_{ \boldsymbol{q}, \lambda}(z, z')  J_{\boldsymbol{q}, \lambda}(z')   \right\}  }{\mathrm{det} \left( - i \hat{G}^{\mathrm{fp}-1}_{\boldsymbol{q}, \lambda} \right) }~,
\label{general formula for bosonic integral}
\end{align}
where
\begin{align}
& J_{\boldsymbol{q}, \lambda}(z) =      M_{-\boldsymbol{q},  \lambda}       \rho_{- \boldsymbol{q}}(z) + F_{-\boldsymbol{q},  \lambda}(z)~, \nonumber \\
& \widetilde{J}_{\boldsymbol{q}, \lambda}(z) =     M_{\boldsymbol{q},  \lambda}       \rho_{ \boldsymbol{q}}(z) + F_{\boldsymbol{q}~,  \lambda}(z)~.
\label{A, J}
\end{align}
In the last step of Eq.~\eqref{general formula for bosonic integral} we have applied to the $\gamma$ contour the standard rules of bosonic Gaussian integration on a continuous time domain, and we have used the direct free-phonon GF given by Eq.~\eqref{Gfp on the contour}. The functional determinant appearing in Eq.~\eqref{general formula for bosonic integral} is given by~\cite{SecchiPolini1, Kamenev}
\begin{align}
& \mathrm{det} \left( - i \hat{G}^{\mathrm{fp}-1}_{\boldsymbol{q}, \lambda} \right)  =   1 -   \mathrm{e}^{- \beta \omega_{\boldsymbol{q} , \lambda}} ~.
\label{determinant A}
\end{align}
By using Eqs.~\eqref{general formula for bosonic integral} and~\eqref{determinant A}, and the definitions in \eqref{A, J}, we find
\begin{align}
 &  \int \mathcal{D} \left( a^*, a \right) \mathrm{e}^{i S_{\mathrm{ep}}  }  \nonumber \\
 &  = \mathrm{Tr} \left(  \mathrm{e}^{- \beta \hat{\cal H}_{\rm p}}  \right)  \exp\Bigg\{ - i \sum_{\boldsymbol{q}, \lambda}  \iint_{\gamma}   \mathrm{d} z \mathrm{d} z' \nonumber \\
& \quad \times \Bigg[        M_{\boldsymbol{q} , \lambda}   \left[ G^{\mathrm{fp}}_{  \boldsymbol{q}, \lambda}\left(z, z' \right)        +     G^{\mathrm{fp}}_{ - \boldsymbol{q}, \lambda}\left(z', z \right)  \right]  F_{-\boldsymbol{q},  \lambda}(z' )           \rho_{\boldsymbol{q}}(z)  \nonumber \\
& \quad +       \left| M_{\boldsymbol{q},  \lambda}  \right|^2   G^{\mathrm{fp}}_{  \boldsymbol{q}, \lambda}\left(z, z' \right)        \rho_{\boldsymbol{q}}(z)   \,       \rho_{-\boldsymbol{q}}(z')  \Bigg]  \Bigg\}~,
\label{bosonic integral done}
\end{align}
where we have used 
\begin{align}
  &     \sum_{\boldsymbol{q} , \lambda} \iint_{\gamma} \mathrm{d} z   \mathrm{d} z'    F_{ \boldsymbol{q}, \lambda}(z)        G^{\mathrm{fp}}_{ \boldsymbol{q}, \lambda}(z, z')   F_{- \boldsymbol{q}, \lambda}(z') = 0~.
\end{align}
The latter can be derived by using Eq.~\eqref{property F on the contour}.

Eq.~\eqref{bosonic integral done} reduces to 
\begin{align}
  Z[ V]     \equiv \frac{\mathrm{Tr}\left(  \mathrm{e}^{- \beta \hat{\cal H}_{\rm p}} \right)}{\mathrm{Tr} \left( \hat{\cal U}_{\gamma_{\rm M}} \right)}
\int \mathcal{D}(\overline{d}, d)\mathrm{e}^{ i S_{\mathrm{eff}}[V]}~,
\label{partition function}
\end{align}
where $S_{\mathrm{eff}}[V]$ is given by Eq.~\eqref{effective action}. In the right-hand side of Eq.~\eqref{partition function} we find the quantity
\begin{align}
\mathrm{Tr} \left( \mathrm{e}^{- \beta \hat{\cal H}_{\rm p}} \right) =   \prod_{\boldsymbol{q}, \lambda} \frac{1}{1 - \mathrm{e}^{- \beta \omega_{{\boldsymbol{q}, \lambda}}}}~.
\end{align}
A further simplification can be obtained by performing the bosonic path integral in the denominator of Eq.~\eqref{partition function}. Since the free-phonon and the phonon-electron interaction Hamiltonians have the same form on the Matsubara branch and on the real-time branches, we have
\begin{align}
\mathrm{Tr} \left( \hat{\cal U}_{\gamma_{\rm M}} \right) & = \mathrm{Tr} \left( \mathrm{e}^{- \beta \hat{\cal H}_{\rm p}} \right) \exp\Bigg\{  i S_{\rm e}^{(\rm{M})}\left[ \overline{d} , d \right] \nonumber \\
& \quad - i \sum_{\boldsymbol{q}, \lambda}  \iint_{\gamma_{\rm M}}   \mathrm{d} z \mathrm{d} z'        \left| M_{\boldsymbol{q},  \lambda}  \right|^2   G^{\mathrm{fp}}_{  \boldsymbol{q}, \lambda}\left(z, z' \right)      \nonumber \\
& \quad \times  \rho_{\boldsymbol{q}}(z)   \,       \rho_{-\boldsymbol{q}}(z')   \Bigg\}    \nonumber \\
& \equiv \mathrm{Tr} \left( \mathrm{e}^{- \beta \hat{\cal H}_{\rm p}} \right) \mathrm{Tr} \left( \hat{\cal{U}}_{\gamma_{\rm M}}^{\rm eff} \right)~,
\end{align}
where $S_{\rm e}^{(\rm{M})}\left[ \overline{d} , d \right]$ is the quadratic electronic action on the Matsubara branch, and it should be noted that the time integrations run only on $\gamma_{\rm M}$. We then obtain Eq.~\eqref{partition function only electrons}.

\section{Hubbard-Stratonovich decoupling and fermionic integration on the KP time contour}
\label{app: HS}

The Hubbard-Stratonovich transformation is based on the following exact identity:
\begin{align}
&  \exp\left\{ - i \int_{\gamma}   \mathrm{d} z   U \sum_{\boldsymbol{q} }  \overline{\Phi}_{ \boldsymbol{q}}(z) \, \Phi_{ \boldsymbol{q}}(z)         \right\}  \nonumber \\
&  =   c    \int   \mathcal{D}  \left[ \frac{\Delta}{U}, \frac{\Delta^*}{U} \right] \, \exp \left\{        
  i \sum_{\boldsymbol{q}  } \int_{\gamma}   \mathrm{d} z \Bigg[   \overline{\Phi}_{\boldsymbol{q}}(z) \, \Delta_{\boldsymbol{q}}(z) \right. \nonumber \\
  & \quad \left. + \, \Phi_{\boldsymbol{q}}(z) \, \Delta^*_{\boldsymbol{q}}(z)   
  +   \frac{1}{ U  }     \left| \Delta_{\boldsymbol{q}}(z)  \right|^2 \Bigg] \right\}~ .
  \label{HS continuum}
\end{align}
After replacing $S_{\mathrm{int}}$ in Eq.~\eqref{effective action} with Eq.~\eqref{Hubbard action}, we use Eq.~\eqref{HS continuum} to simplify Eq.~\eqref{partition function only electrons} into
\begin{align}
  Z\left[V \right] 
& = 
  \frac{c}{\mathrm{Tr} \left( \hat{\cal U}^{\mathrm{eff}}_{\gamma_{\rm M}} \right)}         \int   \mathcal{D}  \left[ \frac{\Delta}{U}, \frac{\Delta^*}{U} \right] \, \int \mathcal{D} \left[\overline{d} , d \right]  \nonumber \\
  & \quad \times 
\exp\left\{   i \iint_{\gamma}   \mathrm{d}z   \mathrm{d}z'  \sum_{\boldsymbol{k} \boldsymbol{k}'  }  \left( \begin{matrix} \overline{d}_{ \boldsymbol{k},  \uparrow }(z), &  \overline{d}_{  \boldsymbol{k},  \downarrow}(z)  \end{matrix}  \right) \right. \nonumber \\
& \quad \times \hat{\mathbf{G}}^{  -1  }_{\boldsymbol{k}, z ; \boldsymbol{k}', z'}[V] 
         \left(  \begin{matrix} d_{\boldsymbol{k}', \uparrow}(z')  \\   d_{   \boldsymbol{k}', \downarrow }(z') \end{matrix} \right) \nonumber \\ 
& \quad \left.          +  \,
  \frac{ i }{ U  } \sum_{\boldsymbol{q}  } \int_{\gamma}   \mathrm{d} z        \left| \Delta_{\boldsymbol{q}}(z)  \right|^2   \right\}~.
 \label{partition function BCS before Grassmann integration}
\end{align}
In Eq.~(\ref{partition function BCS before Grassmann integration}) we have introduced the inverse BCS electronic GF on $\gamma$ (in the presence of sources), see Eq.~\eqref{G_BCS-1} in the main text. The Dirac deltas $\delta(z,z')$ appearing in Eq.~\eqref{G_BCS-1} are shorthands: they connect the $\overline{d}$ and $d$ Grassmann fields appearing in Eq.~\eqref{partition function BCS before Grassmann integration} whose time arguments are infinitesimally shifted along the contour, i.e., $\overline{d}(z)$ is connected with $d(z - 0)$. This has consequences on the determination of the GFs, see Eq.~\eqref{observables path integral formula 1}. 

We proceed by integrating away the Grassmann variables appearing in Eq.~\eqref{partition function BCS before Grassmann integration}. To this aim, we use the general formula for a discrete-time action
\begin{align}
\int \mathcal{D}\left(\overline{d}, d \right)\exp{\left(- {}^T \overline{d} \cdot \mathbf{X} \cdot d\right)} = \det \mathbf{X} = \mathrm{e}^{\mathrm{tr} \left( \mathrm{ln} \mathbf{X} \right)}~.
\label{det X}
\end{align}
Performing the fermionic Gaussian integration, we obtain Eq.~\eqref{introducing nonequilibrium action Delta} for the Hubbard-BCS partition function $Z[V]$.

	Note that the first term on the right-hand side of Eq.~\eqref{S_BCS} is a formal shorthand: the operator $\hat{G}^{-1}$ should be replaced by its discrete-time~\cite{Kamenev, SecchiPolini1} version~$G^{-1}$. The latter is a matrix defined on discrete contour coordinates (as well as on the other indexes), whose elements we denote by $G^{-1}_{z, z'}$~. If $G_{z, z'}$ is the direct GF on the contour, we have the following properties: i) if the time coordinates are taken on a discrete grid $\Gamma$, then
\begin{align}
\sum_{z' \in \Gamma} G^{-1}_{z, z'} G_{z', z''} = \delta_{z, z''}~,
\end{align} 
where $\delta_{z, z''} $ is the Kronecker delta; ii) if the contour coordinates are taken on the continuous contour $\gamma$, then
\begin{align}
\int_{\gamma} \mathrm{d} z' \hat{G}^{-1}_{z, z'} G_{z', z''} = \delta\left(z, z''\right)~,
\end{align}
where $\delta\left(z, z''\right)$ is the Dirac delta on the contour. When the discrete-time form is used, the quantity 
\begin{align*}
\mathrm{tr} \left[ \mathrm{ln} \left( - i   G^{ -1}[V]     \right) \right]
\end{align*}
is well-defined, with the understanding that the trace should be taken also with respect to the $z$ coordinates on the grid $\Gamma$. The continuum limit can always be taken at the end of the derivation. This is exactly what we do in Appendix~\ref{app: functional derivatives}---see Eq.~\eqref{continuum limit after functional diff}.

\section{Functional derivatives}
\label{app: functional derivatives}

In the main text, we have used the functional derivatives of Eq.~\eqref{S_BCS} with respect to $V^{\downarrow \uparrow}, V^{\uparrow \downarrow}, \Delta, \Delta^*$. The first step can be done in general. If $x$ is the field with respect to which we differentiate, then
\begin{align}
& \frac{\delta}{\delta x} \mathrm{tr} \left\{ \mathrm{ln} \left( - i    G^{-1}[V]    \right)\right\}    = \mathrm{tr} \left\{     G[V] \frac{\delta}{\delta x}    G^{ -1   }[V]   \right\}  
   \nonumber \\
& =  
\sum_{\boldsymbol{k} ,  \boldsymbol{k}' } \sum_{ \sigma , \sigma' }\iint_{ \gamma} \mathrm{d}z \mathrm{d}z'      G_{\boldsymbol{k}, \sigma, z ; \, \boldsymbol{k}'  , \sigma' , z' }[V]   \frac{\delta}{\delta x}     G^{ -1  }_{\boldsymbol{k}', \sigma' , z' ; \, \boldsymbol{k} , \sigma , z  }[V]  .
\label{continuum limit after functional diff}
\end{align}
From Eq.~\eqref{G_BCS-1}, specifying the fields that we need, we get
\begin{align}
& \frac{\delta \, \mathrm{tr} \left\{ \mathrm{ln} \left( - i    G^{ -1  }[V]   \right)\right\}}{\delta \Delta^*_{\boldsymbol{q}}(z)}     =  
\sum_{\boldsymbol{k}   }           G_{\boldsymbol{k}, \uparrow, z ; \,  \boldsymbol{k} + \boldsymbol{q}   , \downarrow , z + 0 }[V] ~, 
\end{align}
and
\begin{align}
& \frac{\delta \, \mathrm{tr} \left\{ \mathrm{ln} \left( - i  G^{-1}[V]  \right)\right\} }{\delta \Delta_{\boldsymbol{q}}(z)}     =  
\sum_{\boldsymbol{k}  }     G_{\boldsymbol{k}, \downarrow, z ; \, \boldsymbol{k}  - \boldsymbol{q} , \uparrow , z + 0 }[V]  ~, 
\end{align}
\begin{align}
& \frac{\delta \, \mathrm{tr} \left\{ \mathrm{ln} \left( - i  G^{-1  }[V]   \right)\right\}}{\delta V^{  \sigma, - \sigma}_{\boldsymbol{k}, \boldsymbol{k}'}(z)}  = - 
          G_{\boldsymbol{k}', - \sigma , z ; \, \boldsymbol{k} ,  \sigma , z + 0 }[V] ~.
\end{align}

\section{Application of APT to nonequilibrium superconductivity}
\label{app: APT details}
\subsection{Time derivatives of the quasiparticle fields}
\label{app: TD quasiparticle}

The time derivatives of the IQP fields are expressed in terms of the IQP fields themselves as
\begin{align}
& \partial_s \hat{D}_{\boldsymbol{k}, \alpha}(s) = \sum_{\alpha'} A_{\boldsymbol{k}}(s)_{\alpha, \alpha'} \hat{D}_{\boldsymbol{k}, \alpha'}(s) ~, 
\label{in terms of A}
\end{align}
where
\begin{align}
 A_{\boldsymbol{k}}(s)_{\alpha, \alpha'} =   \left[\partial_s {a}^*_{\boldsymbol{k}, \alpha}(s) \right] a_{\boldsymbol{k}, \alpha'}(s) + \left[\partial_s b^*_{\boldsymbol{k}, \alpha}(s) \right] b_{\boldsymbol{k}, \alpha'}(s)~.
 \label{explicit A}
\end{align}
Because of Eq.~\eqref{orthonormalization}, one has
\begin{align}
A_{\boldsymbol{k}}(s)_{\alpha, \alpha'} = - A^*_{\boldsymbol{k}}(s)_{\alpha', \alpha}~. 
\label{symmetry A alpha alpha'}
\end{align}
Using Eqs.~\eqref{explicit eigenstates} and \eqref{complex gap}, we obtain
\begin{align}
 A_{\boldsymbol{k}}(s)_{\alpha, \alpha}   =   \frac{- i  \left| \Delta(s) \right|^2 \partial_s \phi(s) }{2 E_{\boldsymbol{k}}(s)    \left\{ E_{\boldsymbol{k}}(s)   -  \alpha \left[ \epsilon_{\boldsymbol{k}} + f(s)  \right] \right\}     }
 \label{A alpha alpha}
\end{align}
and
\begin{align}
  A_{\boldsymbol{k}}(s)_{- \alpha , \alpha}   \equiv   \frac{\left| \Delta(s) \right|}{2} \left[ \alpha V_{ \boldsymbol{k}}(s) + i W_{ \boldsymbol{k}}(s)  \right]~.
        \label{hopefully small}
\end{align}
Here, the real quantities $V_{ \boldsymbol{k}}(s)$ and $W_{ \boldsymbol{k}}(s)$ are given by
\begin{align}
  V_{ \boldsymbol{k}}(s) & =  \frac{1}{E^2_{ \boldsymbol{k}}(s) }  \left[    \partial_s f(s)          
    -      \frac{ \epsilon_{\boldsymbol{k}} + f(s) }{\left| \Delta(s) \right|} \, \partial_s \left| \Delta(s) \right|  \right]   \nonumber \\
    &    =   \frac{\partial_s f(s) - \left[ \epsilon_{\boldsymbol{k}} + f(s) \right] {\rm Re}\left[  \partial_s  \Delta(s) / \Delta(s) \right] }{  E^2_{ \boldsymbol{k}}(s) }
\label{V adiabaticity}
\end{align}
and
\begin{align}
W_{ \boldsymbol{k}}(s) & =    \frac{    \partial_s \phi(s)     }{  E_{ \boldsymbol{k}}(s)   } =   \frac{  {\rm Im}\left[  \partial_s  \Delta(s) / \Delta(s)  \right] }{ E_{ \boldsymbol{k}}(s)      }~.
\label{W adiabaticity}
\end{align}
\subsection{Geometrical and dynamical phase factors}

Using Eq.~\eqref{instantaneous spectrum} and the results in Appendix \ref{app: TD quasiparticle}, we can calculate the dynamical and geometrical factors. We find
\begin{align}
\omega_n(t) = - \sum_{\boldsymbol{k}, \alpha} \alpha n_{\boldsymbol{k}, \alpha} \int_{s_0}^s \mathrm{d} s'  E_{\boldsymbol{k}}(s')
\label{dynamical phase}
\end{align}
and
\begin{align}
 \gamma_n(s) & = i \sum_{\boldsymbol{k}, \alpha} n_{\boldsymbol{k}, \alpha} \int_{s_0}^{s} \mathrm{d} s' \langle 0_D  | \hat{D}_{\boldsymbol{k}, \alpha}(s') \, \partial_{s'} \hat{D}^{\dagger}_{\boldsymbol{k}, \alpha}(s')  | 0_D \rangle   \nonumber \\
      &    = i \sum_{\boldsymbol{k}, \alpha} n_{\boldsymbol{k}, \alpha} \int_{s_0}^{s} \mathrm{d} s' A^*_{\boldsymbol{k}}(s')_{\alpha, \alpha}    \nonumber \\
&  = \!  \sum_{\boldsymbol{k}, \alpha} n_{\boldsymbol{k}, \alpha} \! \int_{s_0}^{s} \mathrm{d} s'   \frac{- \left| \Delta(s') \right|^2 \partial_{s'} \phi(s') }{2 E_{\boldsymbol{k}}(s')    \left\{ E_{\boldsymbol{k}}(s')   -  \alpha \left[ \epsilon_{\boldsymbol{k}} + f(s')  \right] \right\}}  .
\end{align}
\subsection{Components of the adiabatic parameter}

From Eq.~\eqref{adiabatic factor} at $m \neq n$ and using Eq.~\eqref{eff H diagonalized} we have
\begin{align}
 M_{n, m}(s) 
 &  = \frac{  1 }{\mathcal{E}_{m}(s) - \mathcal{E}_{n}(s)}  \sum_{\boldsymbol{k}, \alpha}  \alpha   \left< n(s) \right| \left\{ \partial_s E_{\boldsymbol{k}}(s)    \hat{N}_{\boldsymbol{k}, \alpha}(s) \right. \nonumber \\
& \quad   +   E_{\boldsymbol{k}}(s)    \left[ \partial_s \hat{D}^{\dagger}_{\boldsymbol{k}, \alpha}(s) \, \hat{D}_{\boldsymbol{k}, \alpha}(s) \right.  \nonumber \\
& \quad \left. \left. + \hat{D}^{\dagger}_{\boldsymbol{k}, \alpha}(s) \partial_s \hat{D}_{\boldsymbol{k}, \alpha}(s) \right]   \right\}   \left| m(s) \right>~,
\label{adiabatic parameter 2}
\end{align}
where $\hat{N}_{\boldsymbol{k}, \alpha}(s) \equiv \hat{D}^{\dagger}_{\boldsymbol{k}, \alpha}(s)  \hat{D}_{\boldsymbol{k}, \alpha}(s)$. We now consider the quantity in curly brackets on the right-hand side of Eq.~(\ref{adiabatic parameter 2}). The first term vanishes because
\begin{align}
\left< n(s) \right|     \hat{N}_{\boldsymbol{k}, \alpha}(s) \left| m(s) \right> = n_{\boldsymbol{k}, \alpha} \delta_{n, m} 
\label{useful}
\end{align}
and we are considering only $m \neq n$. To calculate the second and third terms, we use Eqs.~\eqref{in terms of A} and Eq.~\eqref{useful}. We get
\begin{align}
M_{n, m}(s) & =   \frac{ - 2 }{\mathcal{E}_{m}(s) - \mathcal{E}_{n}(s)}  \sum_{\boldsymbol{k}, \alpha} \alpha  E_{\boldsymbol{k}}(s)       A_{\boldsymbol{k}}(s)_{- \alpha,  \alpha}     \nonumber \\
& \quad \times   \left< n(s) \right| \hat{D}^{\dagger}_{\boldsymbol{k}, - \alpha}(s)  \hat{D}_{\boldsymbol{k}, \alpha}(s) \left| m(s) \right>~,
\label{adiabatic parameter 3}
\end{align}
where we have used Eq.~\eqref{symmetry A alpha alpha'}. The bra-ket appearing in the second line of Eq.~\eqref{adiabatic parameter 3} is zero unless the sets $n$ and $m$ are such that $m_{\boldsymbol{k}', \pm 1} = n_{\boldsymbol{k}', \pm 1} \,\, \forall \boldsymbol{k}' \neq \boldsymbol{k}$, while $m_{\boldsymbol{k}, \alpha} = 1 = n_{\boldsymbol{k}, - \alpha}$ and $m_{\boldsymbol{k}, - \alpha} = 0 = n_{ \boldsymbol{k},  \alpha}$. In this case, the bra-ket is equal to $1$. To make the notation compact, when the set $m$ satisfies these conditions with respect to the set $n$, we will write that $m = n[\boldsymbol{k}, \alpha]$. We can then write 
\begin{align}
& \left< n(s) \right|       \hat{D}^{\dagger}_{\boldsymbol{k}, - \alpha}(s) \hat{D}_{\boldsymbol{k}, \alpha}(s) \!  \left| m(s) \right> \!  =   
\delta_{ n_{\boldsymbol{k}, - \alpha}, 1  } \,
\delta_{ n_{\boldsymbol{k},   \alpha}, 0  } \, \delta_{m ,  n[\boldsymbol{k}, \alpha]  } .
\label{useful brackets}
\end{align}
Inserting Eq.~\eqref{useful brackets} into Eq.~\eqref{adiabatic parameter 3}, and observing, with the aid of Eq.~\eqref{instantaneous spectrum}, that $\mathcal{E}_{ m }(s) - \mathcal{E}_{ n}(s) = 2 \alpha  E_{\boldsymbol{k}}(s)$ for all the sets $n$ and $m$ such that Eq.~\eqref{useful brackets} equals 1, we finally find:
\begin{align}
M_{n, m}(s)  & =   - \sum_{\boldsymbol{k}, \alpha}     \delta_{ n_{\boldsymbol{k}, - \alpha}, 1  } \,
\delta_{ n_{\boldsymbol{k},   \alpha}, 0  } \,  \delta_{m ,  n[\boldsymbol{k}, \alpha]  }     A_{\boldsymbol{k}}(s)_{ - \alpha,  \alpha}~.
\label{adiabatic parameter 4}
\end{align}
Inserting this result into Eq.~\eqref{Jmn} and using Eq.~\eqref{hopefully small}, we get (for $m \neq n$)
\begin{align}
 J_{n , m}(s)   &  = - \frac{1}{8}  \sum_{\boldsymbol{k}, \alpha} \alpha \,       \delta_{ n_{\boldsymbol{k}, - \alpha}, 1  } \,
\delta_{ n_{\boldsymbol{k},   \alpha}, 0  } \,  \delta_{m ,  n[\boldsymbol{k}, \alpha]  }      \nonumber \\
   & \quad \times  \int_{s_0}^s \mathrm{d} s'      \frac{ \left| \Delta(s') \right|^2}{ E_{\boldsymbol{k}}(s') } \left[   V^2_{  \boldsymbol{k}}(s') + W^2_{  \boldsymbol{k}}(s')  \right]~.  
   \label{expression J}
\end{align}
\section{Leading-order APT components of the nonequilibrium gap}
\label{app: Gamma}
We here present some details on the derivation of the quantities $\Gamma_{\Delta}^{(0)}(s)$ and $\Gamma_{\Delta}^{(0)}(s)$, which are needed to compute the terms with $p = 0$ and $p = 1$ of Eq.~\eqref{Delta(s) adiabatic expansion}. Their expressions are
\begin{align}
\Gamma_{\Delta}^{(0)}(s) = - U   \sum_{\boldsymbol{k}, \Psi_0}   W_{\Psi_0}    \left\langle \Psi^{(0)}(s)  \right|   \hat{d}^{\dagger}_{\boldsymbol{k}, \downarrow } \hat{d}_{\boldsymbol{k}, \uparrow }       \left| \Psi^{(0)}(s)  \right\rangle
\label{Delta0(s) expression}
\end{align}
and
\begin{align}
\Gamma_{\Delta}^{(1)}(s) & =  - U   \sum_{\boldsymbol{k}} \sum_{\Psi_0}   W_{\Psi_0} \Big\{ \left\langle \Psi^{(1)}(s) \right| \hat{d}^{\dagger}_{\boldsymbol{k}, \downarrow } \hat{d}_{\boldsymbol{k}, \uparrow }    \left| \Psi^{(0)}(s) \right\rangle 
\nonumber \\ & \quad   + \left\langle \Psi^{(0)}(s) \right| \hat{d}^{\dagger}_{\boldsymbol{k}, \downarrow } \hat{d}_{\boldsymbol{k}, \uparrow }    \left| \Psi^{(1)}(s) \right\rangle \Big\}~,
\label{Delta1(s) expression}
\end{align}
respectively.

In order to obtain explicit expressions, we take into account Eqs.~\eqref{Psi0 final} and~\eqref{Psi1 final} and combine the results of Sect.~\ref{subsect: Instantaneous} with those of Appendix~\ref{app: APT details}. We also note that
\begin{align}
\hat{d}^{\dagger}_{\boldsymbol{k}, \downarrow } \hat{d}_{\boldsymbol{k}, \uparrow } = 
\sum_{\alpha, \alpha'} b^*_{\boldsymbol{k}, \alpha}(t) a_{\boldsymbol{k}, \alpha'}(t)
     \hat{D}^{\dagger}_{\boldsymbol{k}, \alpha}(t)    \hat{D}_{\boldsymbol{k}, \alpha'}(t)~,
\label{dd DD}
\end{align}
and we use Eqs.~\eqref{explicit eigenstates}.

After some algebra, we find
\begin{align}
\Gamma^{(0)}_{\Delta}(s) & = -   \Delta(s)   \frac{U}{2}   \sum_{\boldsymbol{k}}   \frac{ w_{\boldsymbol{k}} }{ E_{\Delta, \boldsymbol{k}}(s)        }    ~,
\label{Gamma0(s) result}
\end{align}
where $w_{\boldsymbol{k}}$ is given in Eq.~\eqref{w_k}, and
\begin{widetext}
\begin{align}
\Gamma^{(1)}_{\Delta}(s) 
  = &  \,   \frac{ U  }{4}       \sum_{\boldsymbol{k}} w_{\boldsymbol{k}}  \left\{  i \, \frac{\Delta(s)  \partial_s f(s) - \left[ \epsilon_{\boldsymbol{k}} + f(s) \right]     \partial_s \Delta(s)   }{  E^3_{\Delta, \boldsymbol{k}}(s) }     \right. \nonumber \\  
          & + \Delta(s) \left[ i   \sin\left[ \theta_{\Delta, \boldsymbol{k}}(s) \right]   - \frac{     \epsilon_{\boldsymbol{k}} + f(s)   }{   E_{\Delta, \boldsymbol{k}}(s)     } \cos\left[ \theta_{\Delta, \boldsymbol{k}}(s) \right]  \right]    \frac{ {\rm Im}\left[  \partial_s{\Delta}(s_0) / \Delta(s_0)  \right] }{  E^2_{\Delta, \boldsymbol{k}}(s_0)      }    \nonumber \\
          & \left.   - \Delta(s)   \left[    i   \cos\!\left[ \theta_{\Delta, \boldsymbol{k}}(s) \right]  + \frac{     \epsilon_{\boldsymbol{k}} + f(s)   }{   E_{\Delta, \boldsymbol{k}}(s)     }  \sin\! \left[ \theta_{\Delta, \boldsymbol{k}}(s) \right]  \right]  \!  \frac{\partial_s f(s_0) - \left[ \epsilon_{\boldsymbol{k}} + f(s_0) \right] {\rm Re} \! \left[  \partial_s \Delta(s_0) / \Delta(s_0) \right] }{  E^3_{\Delta, \boldsymbol{k}}(s_0) } 
           \right\}  ~,     
\label{Gamma1(s) result}
\end{align}
\end{widetext}
where we have used Eqs.~\eqref{V adiabaticity} and \eqref{W adiabaticity}, as well as the shorthand Eq.~\eqref{alpha theta}.

From Eq.~\eqref{Gamma0(s) result} onwards, we have explicitly indicated which quantities depend on $\Delta$, e.g.~by writing $E_{\boldsymbol{k}}(s) \rightarrow E_{\Delta, \boldsymbol{k}}(s)$. Because $\theta_{\Delta, \boldsymbol{k}}(s_0) = 0$ (see Eq.~\eqref{alpha theta}), one can verify that $\Gamma^{(1)}_{\Delta}(s_0) = 0$.

We now put in correspondence the two expansions given by Eqs.~\eqref{Delta(s) adiabatic expansion} and~\eqref{equation for Delta(s) adiabatic expansion}. In order to identify $\Delta^{(0)}(s)$ and $\Delta^{(1)}(s)$ we put $\Delta(s) \approx \Delta^{(0)}(s) + T^{-1} \Delta^{(1)}(s)$ in Eq.~\eqref{Delta0(s) expression} and we expand by assuming that the second term is small, obtaining
\begin{align}
\Gamma^{(0)}_{\Delta}(s) & \approx -   \Delta^{(0)}(s)   \frac{U}{2}   \sum_{\boldsymbol{k}}   \frac{ w_{\boldsymbol{k}} }{ E_{\Delta^{(0)}, \boldsymbol{k}}(s)    }     \nonumber \\
& \quad -  \frac{1}{T} \, \Delta^{(1)}(s) X_{ \Delta^{(0)}}(s)~,
\label{Gamma0(s) expansion in Delta}
\end{align}
where $E_{\Delta^{(0)}, \boldsymbol{k}}(s)   \equiv \sqrt{\left[ \epsilon_{\boldsymbol{k}}    +   f(s)  \right]^2 + \left| \Delta^{(0)}(s) \right|^2   }$ and we have introduced the quantity in Eq.~\eqref{X_Delta}. The first line in the right-hand side of Eq.~\eqref{Gamma0(s) expansion in Delta} should then be identified with $\Delta^{(0)}(s)$, while the second line contributes to the term $T^{-1} \Delta^{(1)}(s)$. The other contribution to the latter is obtained from Eq.~\eqref{Gamma1(s) result} evaluated at $\Delta \rightarrow \Delta^{(0)}$.

\section{Simplifications of the validity condition}
\label{app: simplifications validity}

	The quantity in Eq.~\eqref{condition APT-1 computational}, which can be used to assess the validity of first-order APT for a specific system, can be simplified as following. Since $\Delta^{(0)}(t)$ can be chosen as real (as discussed in the main text), Eq.~\eqref{hopefully small} evaluated at $\Delta \rightarrow \Delta^{(0)}$ reduces to 
\begin{align} 
A_{\boldsymbol{k}}(s)_{- \alpha , \alpha}   =  \alpha  \left| \Delta^{(0)}(s) \right|     V_{ \Delta^{(0)}, \boldsymbol{k}}(s)/2~.
\end{align} 
Then, Eq.~\eqref{V adiabaticity} (see the first line) requires $\partial_s \left| \Delta^{(0)}(s) \right| $. It is convenient to write this quantity in terms of $\Delta^{(0)}(s)$, $f(s)$, and $\partial_s f(s)$. This can be done by taking the derivative with respect to $s$ of Eq.~\eqref{Equation Delta0 supercond}, which yields
\begin{align}\label{derivative Delta - derivative f}
| \Delta^{(0)}(s)| \partial_s | \Delta^{(0)}(s)|  \equiv - [ f(s) + J_{\Delta^{(0)}}(s)]  \partial_s f(s)~,
\end{align}
where $J_{\Delta^{(0)}}(s)$ is given by Eq.~\eqref{J}. Note that Eq.~\eqref{J} is a weighted sum, constrained by $ E_1  <  J_{\Delta^{(0)}}(s) <  E_2 ~\forall s$. Finally, Eq.~\eqref{derivative Delta - derivative f} can be used to manipulate Eq.~\eqref{condition APT-1 computational} only if $  \Delta^{(0)}(s)   \neq 0$; otherwise it gives us information on that $\Delta^{(0)}(s)$ can vanish instantaneously at $s$ (while being allowed to be non-zero at other times) only if $[ f(s) + J_{\Delta^{(0)}}(s)]\partial_s f(s) = 0$. For the sake of simplicity, we assume that we are in a situation in which $\Delta^{(0)}(s)   \neq 0$. Then, after some straightforward algebraic manipulations, we obtain Eq.~\eqref{after algebra}.

\section{Derivation of the first-order APT component of the nonequilibrium gap at zero temperature}
\label{app: simplifications Delta_1}
By using the observations made at the beginning of Section \ref{subsec: ex Delta1}, Eq.~\eqref{Equation Delta1} is significantly simplified into
\begin{align}
  \Delta^{(1)}(t)     & =    \frac{ i U  }{4 \left[ 1 + X_{ \Delta^{(0)}}(t) \right] }    \nonumber \\
  & \quad \times   \sum_{\boldsymbol{k}}       \frac{\Delta^{(0)}(t)  \dot{f}(t) -   \left[  f(t) + \epsilon_{\boldsymbol{k}}   \right] \dot{\Delta}^{(0)}(t)     }{  \left\{  \left[ \epsilon_{\mathbf{k}} + f(t) \right]^2 + \left| \Delta^{(0)}(t) \right|^2 \right\}^{3/2}  }~.
           \label{Equation Delta1 under simplification} 
\end{align}
To proceed, we need several ingredients. We start with $X_{ \Delta^{(0)}}(t)$, which is given by Eq.~\eqref{X_Delta} with $\phi^{(0)}(t) = 0$ because $\Delta^{(0)}(t) \geq 0$. We take $w_{\bm{k}} = 1$ and we use Eq.~\eqref{what we want} to obtain
\begin{align}
X_{ \Delta^{(0)}}(t) & = - 1 -     \frac{U}{2}   \sum_{\boldsymbol{k}}      \frac{ \left[ \Delta^{(0)}(t)  \right]^2 }{    \left\{ \left[ \epsilon_{\mathbf{k}} + f(t) \right]^2 + \left| \Delta^{(0)}(t) \right|^2      \right\}^{3/2} }~,
\end{align}
and the prefactor in the right-hand side of Eq.~\eqref{Equation Delta1 under simplification} is simplified as
\begin{align}
& \frac{ i U  }{4 \left[ 1 + X_{ \Delta^{(0)}}(t) \right] } \nonumber \\
&  = - \frac{ i }{2   \left[ \Delta^{(0)}(t)  \right]^2  \sum_{\boldsymbol{k}}        \left\{ \left[ \epsilon_{\mathbf{k}} + f(t) \right]^2 + \left| \Delta^{(0)}(t) \right|^2      \right\}^{- 3/2}    }~.
\end{align}
In order to simplify the first line of Eq.~\eqref{Equation Delta1 under simplification}, we then need to use Eq.~\eqref{conversion k energy} twice. First, we compute the integral
\begin{align}
& \sum_{\boldsymbol{k}}        \frac{1}{ \left\{\left[ \epsilon_{\mathbf{k}} + f(t) \right]^2 + \left| \Delta^{(0)}(t) \right|^2      \right\}^{3/2}  } \nonumber \\
& = \sigma_0 \int_{E_1}^{E_2}   d\epsilon \frac{1}{ \left\{\left[ \epsilon  + f(t) \right]^2 + \left| \Delta^{(0)}(t) \right|^2      \right\}^{3/2}  } \nonumber \\
& = \sigma_0 \frac{1}{\left| \Delta^{(0)}(t) \right|^2} 
\left( \frac{f(t) + E_2}{  R_2(t) }  - \frac{f(t) + E_1}{  R_1(t) }   \right)~,
\label{integral 1}
\end{align}
where we have introduced, for convenience, the quantity $R_i(t) \equiv \sqrt{ \left[ f(t) + E_i \right]^2 + \left| \Delta^{(0)}(t) \right|^2 }$ ($i = 1, 2$). Then, we compute the integral
\begin{align}
& \sum_{\boldsymbol{k}}        \frac{\epsilon_{\mathbf{k}}}{ \left\{\left[ \epsilon_{\mathbf{k}} + f(t) \right]^2 + \left| \Delta^{(0)}(t) \right|^2      \right\}^{3/2}  } \nonumber \\
& = \sigma_0 \int_{E_1}^{E_2}   d\epsilon \frac{\epsilon}{ \left\{\left[ \epsilon  + f(t) \right]^2 + \left| \Delta^{(0)}(t) \right|^2      \right\}^{3/2}  } \nonumber \\
& = - \sigma_0 \frac{f(t)}{\left| \Delta^{(0)}(t) \right|^2} 
\left( \frac{  f(t)   + E_2}{  R_2(t) }     - \frac{  f(t)   +  E_1}{  R_1(t) }   \right) \nonumber \\
& \quad  - \sigma_0  \left( \frac{    1  }{  R_2(t) }     - \frac{    1  }{  R_1(t) }   \right)~.
\label{integral 2}
\end{align}
The quantities $R_i(t)$, just introduced, can be simplified. Defining $f_C(t) \equiv f(t) - C$, we write for $R_1(t)$
\begin{align}
R_1(t) & = \sqrt{ \left[ f_C(t) - E_{\rm D} \right]^2 + \left| \Delta^{(0)}(t) \right|^2 } \nonumber \\
& = \sqrt{   \left( \frac{  x+1   }{ x-1 } \right)^2 E_{\rm D}^2  + \left( \frac{  x-1     }{ x+1 }  \right)^2 f^2_C(t) - 2 f_C(t) E_{\rm D} } \nonumber \\
& = \left|   \frac{  x+1   }{ x-1 }   E_{\rm D}   -   \frac{  x-1     }{ x+1 }    f_C(t) \right|~,
\label{R1 before absolute value}
\end{align}
where we have used the explicit expression for the zero-order gap in the form of Eq.~\eqref{obvious maximization}. The quantity between the absolute sign symbols in Eq.~\eqref{R1 before absolute value} is positive definite if
\begin{align}
f_C(t) < \left( \frac{  x+1   }{ x-1 } \right)^2   E_{\rm D}~.       
\end{align}
We note that
\begin{align}
\left( \frac{   x+1    }{ x-1 } \right)^2 - \frac{   x+1    }{ x-1 } \equiv y = 2 \frac{   x+1    }{ ( x-1 )^2 }  > 0~.
\end{align}
Therefore the quantity of interest is positive definite if
\begin{align}
f (t) < f_{\rm max}  + y E_{\rm D}~,       
\end{align}
where $f_{\rm max}$ is the maximum value of $f(t)$ allowing for the nonequilibrium gap to exist, see Eq.~\eqref{constraint on f}. Therefore, since $y>0$ and $E_{\rm D} > 0$, we conclude that for every $f(t)$ such that the gap exists, the quantity between absolute value signs in Eq.~\eqref{R1 before absolute value} is strictly positive. Analogously, for $R_1(t)$ we write
\begin{align}
R_2(t) & = \sqrt{ \left[ f_C(t) + E_{\rm D} \right]^2 + \left| \Delta^{(0)}(t) \right|^2 } \nonumber \\
& = \sqrt{   \left( \frac{  x+1   }{ x-1 } \right)^2 E_{\rm D}^2  + \left( \frac{  x-1     }{ x+1 }  \right)^2 f^2_C(t) + 2 f_C(t) E_{\rm D} } \nonumber \\
& = \left|   \frac{  x+1   }{ x-1 }   E_{\rm D}   +   \frac{  x-1     }{ x+1 }    f_C(t) \right|~.
\label{R2 before absolute value}
\end{align}
The quantity between absolute sign values in Eq.~\eqref{R2 before absolute value} is positive definite if
\begin{align}
      f(t) >  C - \left( \frac{  x+1   }{ x-1 } \right)^2  E_{\rm D}  =  f_{\rm min} - y E_{\rm D}~,
\end{align}
where $f_{\rm min}$ is the minimum value of $f(t)$ allowing for the nonequilibrium gap to exist, see Eq.~\eqref{constraint on f}. Therefore, analogously to the previous case, we conclude that the quantity between absolute value signs in Eq.~\eqref{R2 before absolute value} is strictly positive. The simplified expressions are
\begin{align}
& R_j(t)   =     \frac{  x+1   }{ x-1 }   E_{\rm D}   + (-1)^j   \frac{  x-1     }{ x+1 }    f_C(t)~, \quad j = 1 , 2~.
\label{R1 and R2 simplified}
\end{align}
We can then easily compute several quantities that appear in the algebraic steps that allow to simplify Eq.~\eqref{Equation Delta1 under simplification}, namely
\begin{align}
  \frac{ R_1(t)  + R_2(t)}{  R_1(t)  - R_2(t)}    = - \frac{E_{\rm D}  }{  f_C(t)   }          \frac{  (x+1)^2 }{(x-1)^2}~;
\label{absolute value done}
\end{align} 
\begin{align}
& \frac{f(t) + E_2}{  R_2(t) }  - \frac{f(t) + E_1}{  R_1(t) } \nonumber \\
         & =   \frac{ x^2 - 1   }{2 x} \frac{  \left[ \Delta^{(0)}(t) \right]^2  }{\displaystyle   E_{\rm D}^2 \left( \frac{x+1}{x-1} \right)^2          - f^2_C(t) \left( \frac{x-1}{x+1} \right)^2 }~.
\end{align}
We also notice that
\begin{align}
\dot{\Delta}^{(0)}(t) =  - \frac{4 x}{(x+1)^2} \frac{\dot{f}(t)  f_C(t) }{  \Delta^{(0)}(t)} ~.        
\end{align}
Combining the relations above and carrying out some straightforward algebraic manipulations, we simplify Eq.~\eqref{Equation Delta1 under simplification} into Eq.~\eqref{Equation Delta1 simplified}.


\begin{thebibliography}{50}
%



\bibitem{Tinkham} 
M. Tinkham, {\it Introduction to Superconductivity}, 2nd ed. (McGraw Hill, Inc., 1996).

\bibitem{Anderson13} 
P.W. Anderson, \href{http://dx.doi.org/10.1088/1742-6596/449/1/012001}{J. Phys.: Conf. Ser.~{\bf 449}, 012001 (2013)}.

\bibitem{Keimer15} 
B. Keimer, S.A. Kivelson, M.R. Norman, S. Uchida, and J. Zaanen, 
\href{http://dx.doi.org/10.1038/nature14165}{Nature~{\bf 518}, 179 (2015)}.


\bibitem{Pietronero95} 
L. Pietronero, S. Str\"assler, and C. Grimaldi, \href{http://dx.doi.org/10.1103/PhysRevB.52.10516}{Phys. Rev. B~{\bf 52}, 10516 (1995)}. 

\bibitem{Grimaldi95} 
C. Grimaldi, L. Pietronero, and S. Str\"assler, \href{http://dx.doi.org/10.1103/PhysRevB.52.10530}{Phys. Rev. B~{\bf 52}, 10530 (1995)}. 

\bibitem{Fausti11} 
D. Fausti, R. Tobey, N. Dean, S. Kaiser, A. Dienst, M. Hoffmann, S. Pyon, T. Takayama, H. Takagi, and A. Cavalleri, \href{http://dx.doi.org/10.1126/science.1197294}{Science~ {\bf 331}, 189 (2011)}.

\bibitem{Hu14} 
W. Hu, S. Kaiser, D. Nicoletti, 
C.R. Hunt, I. Gierz, M.C. Hoffmann, 
M. Le Tacon, T. Loew, B. Keimer, and A. Cavalleri, \href{http://dx.doi.org/10.1038/nmat3963}{Nature Mater.~{\bf 13}, 705 (2014)}.

\bibitem{Kaiser14} 
S. Kaiser, C.R. Hunt, D. Nicoletti, W. Hu, 
I. Gierz, H. Y. Liu, M. Le Tacon, T. Loew, D. Haug, B. Keimer, and A. Cavalleri, \href{http://dx.doi.org/10.1103/PhysRevB.89.184516}{Phys. Rev. B~{\bf 89}, 184516 (2014)}.

\bibitem{mitrano_nature_2016}
M. Mitrano, A. Cantaluppi, D. Nicoletti, 
S. Kaiser, A. Perucchi, S. Lupi, P. Di Pietro, D. Pontiroli, M. Ricc\`{o}, 
S.R. Clark, 
D. Jaksch, and A. Cavalleri, \href{http://dx.doi.org/10.1038/nature16522}{Nature~{\bf 530}, 461 (2016)}.


\bibitem{RammerSmith} 
J. Rammer and H. Smith, 
\href{http://dx.doi.org/10.1103/RevModPhys.58.323}{Rev. Mod. Phys.~{\bf 58}, 323 (1986)}.

\bibitem{Kopnin} 
N.B. Kopnin, {\it Theory of Nonequilibrium Superconductivity} 
(Clarendon Press, Oxford, 2001).

\bibitem{Usadel} 
K. Usadel, 
\href{http://dx.doi.org/10.1103/PhysRevLett.25.507}{Phys. Rev. Lett.~{\bf 25}, 507 (1970)}.

\bibitem{Kamenev} 
A. Kamenev, {\it Field Theory of Non-Equilibrium Systems} (Cambridge University Press, Cambridge, 2011).

\bibitem{Semenov16} A. V. Semenov, I. A. Devyatov, P. J. de Visser, and T. M. Klapwijk, \href{https://doi.org/10.1103/PhysRevLett.117.047002}{Phys. Rev. Lett.~{\bf 117}, 047002 (2016)}.


\bibitem{Abrahams66} 
E. Abrahams and T. Tsuneto, 
\href{http://dx.doi.org/10.1103/PhysRev.152.416}{Phys. Rev.~{\bf 152}, 416 (1966)}.

\bibitem{Messiah}
A. Messiah, {\it Quantum Mechanics, vol. 2,} (John Wiley \& Sons, New York, 1958).

\bibitem{Barankov04} 
R.A. Barankov, L.S. Levitov, and B.Z. Spivak, 
\href{http://dx.doi.org/10.1103/10.1103/PhysRevLett.93.160401}{Phys. Rev. Lett.~{\bf 93}, 160401 (2004)}.

\bibitem{Warner05} 
G.L. Warner and A.J. Leggett, 
\href{http://dx.doi.org/10.1103/PhysRevB.71.134514}{Phys. Rev. B~{\bf 71}, 134514 (2005)}.

\bibitem{Volkov74} 
A.F. Volkov and S.M. Kogan, \href{http://www.jetp.ac.ru/cgi-bin/dn/e_038_05_1018.pdf}{Sov. Phys. JETP~{\bf 38}, 1018 (1974)}.


\bibitem{Barankov06} 
R.A. Barankov and L.S. Levitov, 
\href{http://dx.doi.org/10.1103/PhysRevLett.96.230403}{Phys. Rev. Lett.~{\bf 96}, 230403 (2006)}.

\bibitem{Yuzbashian06} 
E.A. Yuzbashyan and M. Dzero, 
\href{http://dx.doi.org/10.1103/PhysRevLett.96.230404}{Phys. Rev. Lett.~{\bf 96}, 230404 (2006)}.

\bibitem{Tomadin08} 
A. Tomadin, M. Polini, M.P. Tosi, and R. Fazio, 
\href{http://dx.doi.org/10.1103/PhysRevA.77.033605}{Phys. Rev. A~{\bf 77}, 033605 (2008)}.

\bibitem{Peronaci15} 
F. Peronaci, M. Schir\'o, and M. Capone, 
\href{http://dx.doi.org/10.1103/PhysRevLett.115.257001}{Phys. Rev. Lett.~{\bf 115}, 257001 (2015)}.

\bibitem{Papenkort07} 
T. Papenkort, V.M. Axt, and T. Kuhn, 
\href{http://dx.doi.org/10.1103/PhysRevB.76.224522}{Phys. Rev. B~{\bf 76}, 224522 (2007)}.

\bibitem{Unterhinninghofen08} 
J. Unterhinninghofen, D. Manske, and A. Knorr, 
\href{http://dx.doi.org/10.1103/PhysRevB.77.180509}{Phys. Rev. B~{\bf 77}, 180509(R) (2008)}.

\bibitem{Krull14} 
H. Krull, D. Manske, G.S. Uhrig, and A.P. Schnyder, 
\href{http://dx.doi.org/10.1103/PhysRevB.90.014515}{Phys. Rev. B~{\bf 90}, 014515 (2014)}.

\bibitem{Yuzbashyan05} 
E.A. Yuzbashyan, B.L. Altshuler, V.B. Kuznetsov, and V.Z. Enolskii, 
\href{http://dx.doi.org/10.1103/PhysRevB.72.220503}{Phys. Rev. B~{\bf 72}, 220503 (2005)}.
 
\bibitem{Mansart13} 
B. Mansart, J. Lorenzana, A. Mann, A. Odeh, M. Scarongella, M. Chergui, and F. Carbone, \href{http://dx.doi.org/10.1073/pnas.1218742110}{Proc. Natl. Acad. Sci. (USA)~{\bf 110}, 4539 (2013)}.

\bibitem{Matsunaga14} 
R. Matsunaga, N. Tsuji, H. Fujita, A. Sugioka, K. Makise, Y. Uzawa, 
H. Terai, Z. Wang, H. Aoki, and R. Shimano, 
\href{http://dx.doi.org/10.1126/science.1254697}{Science~{\bf 345}, 1145 (2014)}.

\bibitem{Coulthard16} 
J. Coulthard, S.R. Clark, S. Al-Assam, A. Cavalleri, and D. Jaksch, 
\href{http://dx.doi.org/10.1103/PhysRevB.96.085104}{Phys. Rev. B~{\bf 96}, 085104 (2017)}.

\bibitem{Knap16} 
M. Knap, M. Babadi, G. Refael, I. Martin, and E. Demler, 
\href{http://dx.doi.org/10.1103/PhysRevB.94.214504}{Phys. Rev. B~{\bf 94}, 214504 (2016)}.

\bibitem{Bukov16} 
M. Bukov, M. Kolodrubetz, and A. Polkovnikov, 
\href{http://dx.doi.org/10.1103/PhysRevLett.116.125301}{Phys. Rev. Lett.~{\bf 116}, 125301 (2016)}.

\bibitem{Kuwahara16} T. Kuwahara, T. Mori, and K. Saito, \href{http://dx.doi.org/10.1016/j.aop.2016.01.012}{Annals of Physics {\bf{367,}} 96-124 (2016).}


\bibitem{Sentef16} 
M.A. Sentef, A.F. Kemper, A. Georges, and C. Kollath, 
\href{http://dx.doi.org/10.1103/PhysRevB.93.144506}{Phys. Rev. B~{\bf 93}, 144506 (2016)}.

\bibitem{Tsuji15} 
N. Tsuji and H. Aoki, 
\href{http://dx.doi.org/10.1103/PhysRevB.92.064508}{Phys. Rev. B~{\bf 92}, 064508 (2015)}.

\bibitem{Anderson58}
P. W. Anderson, 
\href{http://dx.doi.org/10.1103/PhysRev.112.1900}{Phys. Rev.~{\bf 112}, 1900 (1958)}.


\bibitem{Rigolin08} 
G. Rigolin, G. Ortiz, and V.H. Ponce, 
\href{http://dx.doi.org/10.1103/PhysRevA.78.052508}{Phys. Rev. A~{\bf 78}, 052508 (2008)}.


\bibitem{Schmid66} 
A. Schmid, 
\href{http://dx.doi.org/10.1007/BF02422669}{Phys. Kondens. Mater.~{\bf 5}, 302 (1966)}.

\bibitem{Gorkov68} 
L.P. Gor'kov and G.M. Eliashberg, {Sov. Phys. JETP~{\bf 27}, 328 (1968)}.

\bibitem{SecchiPolini1} A. Secchi and M. Polini, 
\href{https://arxiv.org/abs/1704.01392}{arXiv:1704.01392v1}.

\bibitem{Gersch08} 
R. Gersch, C. Honerkamp, and W. Metzner, 
\href{http://dx.doi.org/10.1088/1367-2630/10/4/045003}{New J. Phys.~{\bf 10}, 045003 (2008)}.

\bibitem{Moulopoulos99} 
K. Moulopoulos and N.W. Ashcroft, 
\href{http://dx.doi.org/10.1103/PhysRevB.59.12309}{Phys. Rev. B~{\bf 59}, 12309 (1999)}. 

\bibitem{Mahan} 
G.D. Mahan, {\it Many Particle Physics} (Kluwer, New York, 2000).

\bibitem{Gorkov58} 
L.P. Gor'kov, \href{http://www.jetp.ac.ru/cgi-bin/dn/e_007_03_0505.pdf}{Sov. Phys. JETP~{\bf 34}, 505 (1958)}.

\bibitem{AltlandSimons} 
A. Altland and B. Simons, {\it Condensed Matter Field Theory} 
(Cambridge University Press, Cambridge, 2010).

\bibitem{Subedi14} 
A. Subedi, A. Cavalleri, and A. Georges, \href{http://dx.doi.org/10.1103/PhysRevB.89.220301}{Phys. Rev. B~{\bf 89}, 220301(R) (2014)}. 

\bibitem{Nicoletti16} 
D. Nicoletti and A. Cavalleri, \href{http://dx.doi.org/10.1364/AOP.8.000401}{Adv. Opt. Photon.~{\bf 8}, 401 (2016)}.

\bibitem{NegeleOrland}
J.W. Negele and H. Orland, {\it Quantum Many-Particle Systems}, (Westview Press, 1998).

\bibitem{Dunnett16}
K. Dunnett and M.H. Szyma\'nska, 
\href{http://dx.doi.org/10.1103/PhysRevB.93.195306}{Phys. Rev. B~{\bf 93}, 195306 (2016)}.

\bibitem{Anghel16} D. Anghel and G.A. Nemnes, \href{https://doi.org/10.1016/j.physa.2016.07.070}{Physica A~{\bf{464,}} 74–82 (2016)}. 


\bibitem{Aronov73} 
A.G. Aronov and V.L. Gurevich, 
\href{http://www.jetp.ac.ru/cgi-bin/dn/e_038_03_0550.pdf}{Sov. Phys. JETP~{\bf 38}, 550 (1974)}.












\end{thebibliography}
\end{document}